\documentclass[journal]{IEEEtran}

\usepackage{cite}		
\usepackage{citesort}
\usepackage[utf8]{inputenc} 
\usepackage[T1]{fontenc}    
\usepackage{soul, color}    
\usepackage{textcomp}	
\usepackage{url}        
\usepackage{microtype}  
\usepackage{flushend}   

\usepackage{amssymb}	
\usepackage{amsfonts}	
\usepackage{amsmath}    
\usepackage{amsthm}     
\usepackage{mathtools}
\usepackage{empheq}     
\usepackage{cases}      

\usepackage{graphicx}	
\usepackage{tabularx}
\usepackage{siunitx}    
\usepackage[caption=false,font=footnotesize]{subfig}
\usepackage[export]{adjustbox}  
\usepackage{booktabs}   

\graphicspath{{Images/}}      
\newcommand{\quotmarks}[1]{``#1''} 
\newcommand{\vect}[1]{\if#1\relax\bm{#1}\else\mathbf{#1}\fi} 
\newtheoremstyle{myremark} 
{}
{}
{}
{}
{\bfseries}
{}
{.5em}
{}

\theoremstyle{myremark} 
\newtheorem*{remark*}{Remark} 

\newcommand\rowheight{1.4}  

\begin{document}

\title{Improved Control Strategies for Intermittent\\ Contact Mode Atomic Force Microscopes}

\author{Marco~Coraggio,
		Martin~Homer,
		Oliver~D.~Payton,
		Mario~di~Bernardo
}

\markboth{IEEE Transactions on Control System Technology,~Vol.~x, No.~x,~September~2016 (Submitted)}%
{Shell \MakeLowercase{\textit{et al.}}: Bare Demo of IEEEtran.cls for IEEE Journals}
  


\maketitle


\begin{abstract}
%

\emph{Atomic force microscopes} have proved to be fundamental research tools in many situations where a gentle imaging process is required, and in a variety of environmental conditions, such as the study of biological samples.
Among the possible modes of operation, \emph{intermittent contact mode} is one that causes less wear to both the sample and the instrument; therefore, it is ideal when imaging soft samples.
However, intermittent contact mode is not particularly fast when compared to other imaging strategies.
In this paper, we introduce three enhanced control approaches, applied at both the dither and z-axis piezos, to address the limitations of existing control schemes.
Our proposed strategies are able to eliminate different image artefacts, automatically adapt scan speed to the sample being scanned and predict its features in real time.
The result is that both the image quality and the scan time are improved.

\end{abstract}

\ifCLASSOPTIONpeerreview
\begin{center}	
\bfseries EDICS Category: 3-BBND
\end{center}
\else
\fi

\IEEEpeerreviewmaketitle


\section{Introduction}
\label{sec:Introduction}

\begin{figure}[t]
 \centering
 \includegraphics[width=\columnwidth]{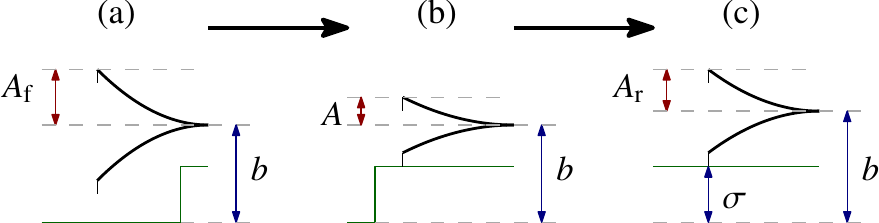}
 \caption{Operation of an intermittent contact mode AFM. (a) Initially, the flexible cantilever (black curved line) oscillates away from the sample (green), with the free oscillation amplitude $A_{\mathrm{f}}$. (b) When the distance between the sample surface and the cantilever is reduced, amplitude decreases to a certain value $A < A_{\mathrm{f}}$. (c) Then, a feedback controller regulates the height $b$ of the base of the cantilever so as the oscillation amplitude $A$ reaches the reference value $A_{\mathrm{r}}$; this way the cantilever is able to perceive the surface, but impacts very gently once every oscillation period.} 
 \label{fig:tm_afm_explanation}
 \end{figure}

\IEEEPARstart{T}{he} \emph{atomic force microscope} (\emph{AFM}) is a device with remarkable precision, used to image hard and soft samples at the nanoscale \cite{and08}, belonging to the family of scanning probe microscopes.
The popularity of the instrument comes from the fact that it can scan samples under ambient temperature and pressure, with a vertical resolution in the order of a hundredth of a nanometre and a lateral resolution in the order of a tenth of a nanometre \cite{fai13}.
The microscope senses sample surfaces by means of a flexible cantilever with an atomically-sharp tip at the end.
When operated in \emph{intermittent contact mode} (\emph{IC-AFM}, also known as \emph{tapping mode}) \cite{fai13}, the cantilever's tip oscillates vertically over the sample surface, driven by a \emph{dither piezo}.
From the way the atomic interaction forces affect the oscillation, it is possible to infer the distance between the sample surface and the tip equilibrium point.
The latter is the position where the tip would be, when at rest; it is also termed the \quotmarks{cantilever base height}, because it corresponds to the height of the fixed end of the cantilever, which is maneuvered by the \emph{\textit{z}-axis piezo}.
As shown in Figure \ref{fig:tm_afm_explanation} (see also Figure \ref{fig:cantilever_geometry}), when far away from the sample, the cantilever oscillates at its maximum (or \emph{free}) oscillation amplitude $A_{\mathrm{f}}$.
When the oscillating cantilever comes close to the sample surface, the interaction forces cause the oscillation amplitude $A$ to decrease; then, a feedback controller, generally a proportional integral derivative (PID) regulator, adjusts the height $b(t)$ of the base of the cantilever so as to attempt to maintain the current oscillation amplitude $A(t)$ at a constant reference value $A_{\mathrm{r}} < A_{\mathrm{f}}$.
The reference amplitude $A_{\mathrm{r}}$ is chosen to balance the need to maximise image quality, while minimising the damage to the AFM tip and sample resulting from impacts.
The height of the sample surface can then be obtained subtracting $A$ from $b$.
At the same time, the sample is moved horizontally under the cantilever, generally in a raster pattern, so as to trace the three-dimensional topography of the sample.
The oscillation amplitude $A(t)$ is extracted in real time from the tip position signal, typically measured using the optical beam deflection method \cite{mey90}, in a process called \emph{demodulation}, operated by a device known as \emph{demodulator}. 

However, although the IC-AFM minimizes damage to the samples while imaging them with great accuracy, the process is hindered by its low speed. 
In this paper we present new control schemes to help address this issue.
Specifically, these strategies allow us to improve image quality by detecting and managing more kinds of image artefacts with respect to established solutions and by predicting features of the samples, exploiting knowledge of those parts which have already been scanned.
Therefore, it is possible to increase scan speed, without worsening image quality.
Furthermore, we propose to adapt scan speed dynamically, depending on the characteristics of the sample, allowing for faster scans, with no effect on imaging accuracy.

The rest of the paper is outlined as follows.
In Section \ref{sec:AFM: a brief overview} we give a detailed explanation of how the IC-AFM works, along with a mathematical formulation.
Then, existing control approaches and their disadvantages are discussed in Section \ref{sec:Existing control approaches}.
After that, in Section \ref{sec:Improved AFM controllers} original solutions are presented to improve the performance and the scanning speed of the microscope.
The novel regulators are validated in Section \ref{sec:Numerical validation} on a set of test samples.
Finally, conclusions are drawn in Section \ref{sec:Conclusions}. 

\section{AFM: a brief overview}
\label{sec:AFM: a brief overview}

 \subsection{Cantilever model}
 \label{subsec:Cantilever model}
 
\begin{figure}[t]
 \centering
 \includegraphics[width=6.5cm]{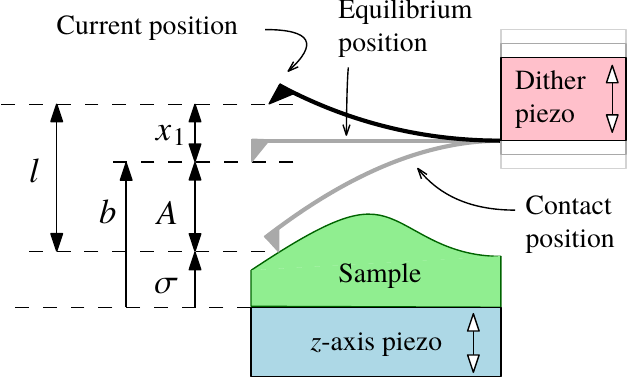}
 \caption{Schematic representation of the cantilever tip close to the sample surface.}
 \label{fig:cantilever_geometry}
 \end{figure}

The cantilever tip is the core of an atomic force microscope, it can be modeled as a mechanical point mass \emph{impact oscillator} \cite{pay11}. 
Specifically, the model can be given as the system
\begin{empheq}[left=\empheqlbrace]{align}\label{eq:hybrid_system_1} 
&\dot{x}_1 = x_2 \\
\label{eq:hybrid_system_2}
&\dot{x}_2 = - \omega_{\mathrm{n}}^2 x_1 - \displaystyle{\frac{\omega_{\mathrm{n}}}{Q}} x_2 + u + F(b + x_1 - \sigma )
\end{empheq}
\begin{equation}\label{eq:dither_piezo}
u = D \sin({\omega_{\mathrm{d}}} t) ,
\end{equation}
when the tip is away from the sample, together with the reset law
\begin{equation}\label{eq:hybrid_system_3}
x_2(t^+) = - r x_2(t^-),
\qquad x_1(t^+) = x_1(t^-) = \sigma(t) - b(t)
\end{equation}
that models the impact between the cantilever tip and the sample surface (in terms of a change in state in the infinitesimally short time before and after an impact, at times $t^-$ and $t^+$ respectively).
In the above equations (see Figure \ref{fig:cantilever_geometry}):
\begin{itemize}
\item $x_1$ is the vertical position [m] of the tip with respect to $b$;
\item $x_2$ is its vertical velocity [m/s];
\item $\omega_{\mathrm{n}} = \sqrt{k/m}$ is the natural (or resonant) frequency [rad/s] of the first flexural mode of the cantilever, with $m$ and $k$ being the mass [kg] and the stiffness coefficient [N/m] of the cantilever, respectively;
\item $Q = m \omega_{\mathrm{n}} /c $ is its quality factor [dimensionless], with $c$ being the damping coefficient [kg/s] of the cantilever;
\item $u$ represents the action of the dither piezo, with $D$ being its driving amplitude [m/s\textsuperscript{2}] and $\omega_{\mathrm{d}}$ its driving frequency [rad/s];
\item $F$ are the interaction forces normalized to mass [m/s\textsuperscript{2}] depending on the distance $l$ [m] between the tip and the sample, which is exactly $l = b + x_1 - \sigma$ (see subsection \ref{subsec:Interaction forces});
\item $b$ is the height [m] of the base of the cantilever;
\item $\sigma$ is the height [m] of the sample surface to be measured;
\item $r$ is the restitution coefficient [dimensionless].
\end{itemize} 
If the cantilever were infinitely far from the the sample, i.e.~assuming $F = 0$ and neglecting (\ref{eq:hybrid_system_3}), at steady state the cantilever tip would oscillate in a sinusoidal motion, with
\begin{equation}
x_1 (t) = A_{\mathrm{f}} \sin(\omega_{\mathrm{d}} t + \varphi),
\end{equation}
where $\varphi$ is a phase shift and the free oscillation amplitude $A_{\mathrm{f}}$ can be computed as 
\begin{equation}\label{eq:A_f}
A_{\mathrm{f}} = \frac{D}{ \left| \omega_{\mathrm{n}}^2 - \omega_{\mathrm{d}}^2 + \cfrac{\omega_{\mathrm{n}}}{Q} i \omega_{\mathrm{d}} \right|} ,
\end{equation}
where $i = \sqrt{-1}$.
In reality, the distance between the cantilever and the sample is finite, therefore $F \neq 0$ and in ideal operation the reset law (\ref{eq:hybrid_system_3}) triggers once every oscillation period, when the tip impacts the sample surface.
As a result, under normal working conditions, with only low velocity impacts, the evolution of tip position in time follows a quasi-sinusoidal motion and can be approximated as
\begin{equation}
x_1 (t) \approx A(t) \sin(\omega_{\mathrm{d}} t + \varphi(t)),
\end{equation}
with $A(t) \approx b(t) - \sigma(t)$ and $A(t) \le A_{\mathrm{f}}$.
Note that this approximation corresponds to the assumption that, since the forcing generated by the dither piezo is close to the fundamental frequency of the cantilever, the amplitude of the fundamental mode is much larger than those of the higher modes, which can be neglected, even though impacts tend to activate all harmonics \cite{ram08}.

 \begin{figure}[t]
 \centering
 \includegraphics[max width=\columnwidth]{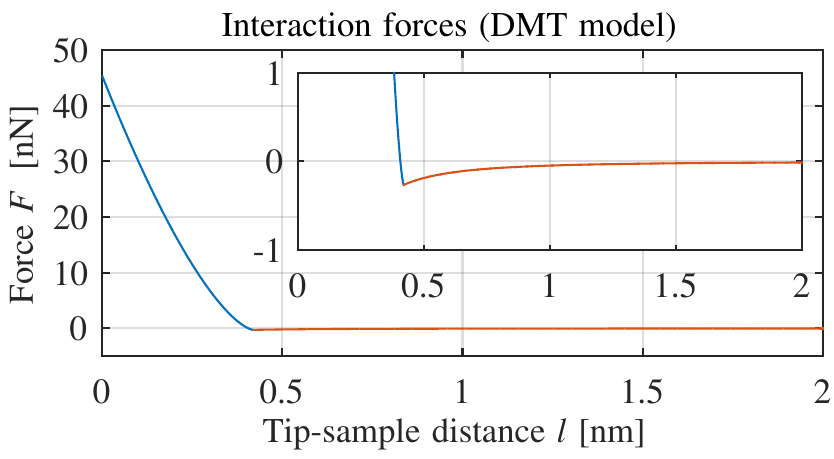}
 \caption{Interaction forces according to Derjaguin-Muller-Toporov model, with coefficient values in Table \ref{tab:afm_parameters}. The inset shows the detail of the repulsive forces.} 
 \label{fig:interaction_forces_dmt}
 \end{figure}
 
\begin{table}[t]
\renewcommand{\arraystretch}{\rowheight}
\caption{Parameters used for the AFM model \cite{pay11}, \cite{kod05}.}
\label{tab:afm_parameters}
\begin{center}
\begin{tabular}{@{}lll@{}}
\toprule
Group  &  Parameter  &  Value  \\
\hline
\emph{Cantilever} &  $\omega_{\mathrm{n}}$  &  $2.85 \cdot 10^{5} \cdot 2 \pi$ rad/s  \\
&  $Q$  &  $100$  \\
&  $r$  &  $0.9$  \\
&  $k$  &  $42$ N/m  \\
&  $m = k/\omega_{\mathrm{n}}^2$  &  $1.3098 \cdot 10^{-11}$ kg  \\
&  $c = m\omega_{\mathrm{n}}/Q$  &  $2.3455 \cdot 10^{-7}$ kg/s  \\
\hline
\emph{Interaction forces}  &  $H$  &  $1.4 \cdot 10^{-19}$ J  \\
&  $r_{\mathrm{t}}$  &  $2$ nm  \\
&  $l_{\mathrm{m}}$  &  $0.42$ nm  \\
&  $E_{\mathrm{t}}, E_{\mathrm{s}}$  &  $1.65 \cdot 10^{11}$ Pa  \\
&  $V_{\mathrm{t}}, V_{\mathrm{s}}$  &  $0.27$  \\
\hline
\emph{\textit{z}-axis piezo}  &  $\omega_{\mathrm{zp}}$  &  $1.5 \cdot 10^6 \cdot 2\pi$ rad/s  \\
&  $Q_{\mathrm{zp}}$  &  $18$  \\
&  $K_{\mathrm{zp}}$  &  $1 / \omega_{\mathrm{zp}}$  \\
\hline
\emph{Feedback controller}  &  $A_{\mathrm{f}}$  &  variable  \\
&  $A_{\mathrm{r}}$  &  $0.9 A_{\mathrm{f}}$  \\  
&  $K_{\mathrm{P}}, K_{\mathrm{D}}$  &  $0$  \\
&  $K_{\mathrm{I}}$  &  $10000$  \\
&  $v_x$  &  $1$ mm/s  \\
\hline
\emph{\textit{Q} control}  &  $Q'$  &  $30$  \\
\hline
\emph{Dynamic/Hybrid PID}  &  $K_{\mathrm{s}}$  &  $15$  \\
&  $\Delta Q_{\mathrm{PL}}, \Delta Q_{\mathrm{RL}}$  &  $25$  \\
&  $A_{\mathrm{t}}^+$  &  $0.95 A_{\mathrm{r}}$  \\
&  $A_{\mathrm{t}}^-$  &  $0.94 A_{\mathrm{r}}$  \\
&  $A_{\mathrm{t,RL}}$  &  $0.5 A_{\mathrm{r}}$  \\
&  $\alpha_{\mathrm{t}}$  &  $-400 A_{\mathrm{f}}$  \\
\hline
\emph{Scan speed regulator}  &  $\tau_v$  &  $0.12$ ms  \\
&  $V_{x,0}$  &  variable  \\
&  $V_{x,\mathrm{m}}$  &  $0.1 V_{x,0}$  \\
&  $V_{x,\mathrm{M}}$  &  $V_{x,0}$  \\
&  $b_{\mathrm{M,a}}$  &  $K_{\mathrm{I}} (A_{\mathrm{r}} - A_{\mathrm{t,RL}})$  \\
&  $b_{\mathrm{M,d}}$  &  $K_{\mathrm{I}} (A_{\mathrm{r}} - A_{\mathrm{t}}^+)$  \\
&  $b_{\mathrm{L,a}}$  &  $0.9 b_{\mathrm{M,a}}$  \\
&  $b_{\mathrm{L,d}}$  &  $0.9 b_{\mathrm{M,d}}$  \\
&  $b_{\mathrm{r,a}}$  &  $0.8 b_{\mathrm{M,a}}$  \\
&  $b_{\mathrm{r,d}}$  &  $0.8 b_{\mathrm{M,d}}$  \\
\hline
\emph{Predictive controller}  &  $M_{\mathrm{PC}}$  &  $3$  \\
&  $E_{\sigma}$  &  $0.1 A_{\mathrm{f}} \cdot I_x$  \\
&  $N_{\mathrm{W}}$  &  $0.01 I_x$  \\
\bottomrule
\end{tabular}
\end{center}
\end{table}

 \subsection{Interaction forces}
 \label{subsec:Interaction forces}
 
There are at least two alternative ways to model the interaction forces $F$ in (\ref{eq:hybrid_system_2}).
The so-called \emph{Lennard-Jones} (\emph{LJ}) \emph{model} \cite{fan05,bas04,sad04} leads to highly stiff force characteristics for small values of the distance $l$ between the tip and the sample, and is therefore discarded here.
Instead we make the approximation that the tip can be modeled as a spherical surface coming in contact with a locally flat sample surface and use the \emph{Derjaguin-Muller-Toporov} (\emph{DMT}) \emph{model} \cite{pay11,mis10,sta10}, where the interaction forces $F$ are given by
\begin{equation}
F(l) = \begin{dcases}
 - \cfrac{H{r_{\mathrm{t}}}}{6{l^2}} ,  &  l > {l_{\mathrm{m}}} \\
 - \cfrac{H r_{\mathrm{t}}}{6l_{\mathrm{m}}^2} + \cfrac{4}{3} \cfrac{ \sqrt{r_{\mathrm{t}} {{({l_{\mathrm{m}}} - l)}^3}}} {  {\cfrac{1 - V_{\mathrm{t}}^2}{E_{\mathrm{t}}} + \cfrac{1 - V_{\mathrm{s}}^2}{E_{\mathrm{t}}}} } ,  &  l \le {l_{\mathrm{m}}}
\end{dcases} ,
\end{equation}
with:
\begin{itemize}
\item $l$ being the tip-sample distance [m];
\item $H$ the Hamaker constant [J];
\item $r_{\mathrm{t}}$ the tip radius [m];
\item $l_{\mathrm{m}}$ the intermolecular distance [m];
\item $E_{\mathrm{t}}$ and $E_{\mathrm{s}}$ the elastic moduli [Pa] of the tip and the sample, respectively;
\item $V_{\mathrm{t}}$ and $V_{\mathrm{s}}$ the Poisson ratios [dimensionless] of the tip and the sample, respectively.
\end{itemize} 
Figure \ref{fig:interaction_forces_dmt} shows the interaction forces as a function of the tip-sample distance $l$ with the parameter values set as in Table \ref{tab:afm_parameters}. 
When the tip and the sample are not too close, there is a small residual attraction between them, due to van der Waals force.
However, when the tip-sample distance is reduced below the intermolecular distance $l_{\mathrm{m}}$ ($l_{\mathrm{m}} \approx 0.42 \text{ nm}$ in Figure \ref{fig:interaction_forces_dmt}), van der Waals force begins to turn repulsive, repulsive Pauli and ionic exclusion forces become prominent and the overall repulsive force becomes larger as $l$ decreases \cite{dan07}.

 \begin{figure}[t]
 \centering
 \includegraphics[max width=\columnwidth]{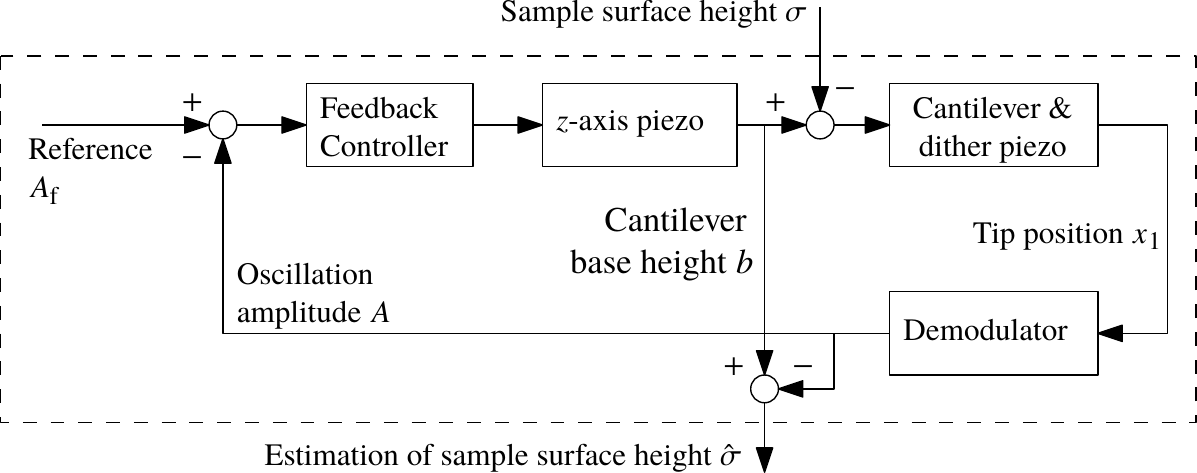}
 \caption{Block diagram representing the intermittent contact mode atomic force microscope.} 
 \label{fig:simplified_tm_afm_scheme_words}
 \end{figure}

 \subsection{Estimation of the sample surface}
 \label{subsec:Estimation of the sample surface}

For correct operation, the oscillation amplitude $A$ must attain a certain constant reference value $A_{\mathrm{r}}$, i.e.
\begin{equation}\label{eq:control_goal}
\mathop {\lim }\limits_{t \to \infty } A(t) = A_{\mathrm{r}}.
\end{equation}
This regulation is fundamental, because, if $A$ becomes too small, the interaction forces will damage the sample; on the other hand, if it becomes too large, the oscillating cantilever tip might lose contact with the sample, causing a phenomenon described in detail in Section \ref{sec:Existing control approaches}, known as \emph{probe loss} or \emph{parachuting}, in which the measurement is incorrect.
Normally, $A_{\mathrm{r}}$ is chosen approximately equal to $0.9 A_{\mathrm{f}}$, with the aim of reducing the magnitude of the interaction forces, whose mean value is proportional to $\sqrt{A_{\mathrm{f}}^2 - A_{\mathrm{r}}^2}$ \cite{fai13}.
In turn, it is common to choose $A_{\mathrm{f}}$ as the smallest value that satisfies $A_{\mathrm{f}} \ge \sigma_{\mathrm{max}} - \sigma_{\mathrm{min}}$, where $\sigma_{\mathrm{max}}$ and $\sigma_{\mathrm{min}}$ are the largest and the smallest values of the sample surface height $\sigma$ on the same scan line (e.g.~\cite{pay11,bas11}).
However, since $\sigma_{\mathrm{max}}$ and $\sigma_{\mathrm{min}}$ are unknown before the scan is performed, $A_\mathrm{f}$ has to be selected conservatively, considering the nature of the sample to be imaged.

To ensure (\ref{eq:control_goal}), a feedback controller is used to adjust $b$, so that an estimate of surface height $\sigma$ can be computed as
\begin{equation}
\hat \sigma = b - A.
\end{equation}
Moreover, at the same time, the sample is moved in a raster pattern on the horizontal \textit{x}-\textit{y} plane, so that the whole specimen is imaged, with the scan lines being parallel to the \textit{x}-axis.
A schematic diagram showing the key components needed for estimating the sample surface height is depicted in Figure  \ref{fig:simplified_tm_afm_scheme_words}.


\section{Existing control approaches}
\label{sec:Existing control approaches}

The problem of controlling $b$ is not straightforward in the framework of control theory.
In fact, it presents a series of complications:
\begin{enumerate}
\item The AFM is a hybrid system because of the reset law (\ref{eq:hybrid_system_3}) due to impacts.
\item The main control input, $b$, affects the behavior of the system through the impact law and also by influencing the interaction forces.
\item While regulating $A$ to $A_{\mathrm{r}}$, the controller has to reject the unknown disturbance due to the sample height $\sigma$.
\item Generally, in the implementation, the relation between $A$ and state vector $[x_1 \; x_2]^\mathrm{T}$ is non-instantaneous. For example, the simplest sample-and-hold demodulators require, at the very least, half an oscillation period to update the value of $A$ \cite{fai13}, thus introducing a delay (see also \cite{abr11}). 
\end{enumerate}
All of these issues make it very complicated to engineer a controller for $b$ and to prove its validity analytically.
Because of this, a relatively simple scheme such as a PID is a well-established solution to the problem \cite{and08}.
Specifically,
\begin{equation}\label{eq:b}
b(t) = \mathrm{PID}(e_A),
\end{equation}
where $e_A(t) = A_{\mathrm{r}} - A(t)$ is the error on the oscillation amplitude and and the PID control action is defined as
\begin{equation}\label{eq:pid}
\mathrm{PID}(\xi(t)) = K_{\mathrm{P}} \xi(t) + K_{\mathrm{I}} \int_0^t \xi(\tau) \, \mathrm{d}\tau + K_{\mathrm{D}} \frac{\mathrm{d} \xi(t)}{\mathrm{d}t} ,
\end{equation}
with $K_{\mathrm{P}}$, $K_{\mathrm{I}}$ and $K_{\mathrm{D}}$ being constant gains.
Nevertheless, since the imaging accuracy given by the PID is not exceptionally good, scan speed cannot be too high.

Moreover, this simple regulator does not implement any mechanism to correctly deal with probe losses, occurring when the sample surface decreases rapidly and the oscillating cantilever loses contact with the sample. 
While there is no contact between the tip and the sample, the measurement is incorrect; therefore this condition is highly undesirable.
Normally, it takes a while to re-establish the contact, because the error on the oscillation amplitude, $e_A$ --- that is the input to the feedback controller regulating $b$ --- eventually saturates to the negative value $A_{\mathrm{r}} - A_{\mathrm{f}}$.

Two regulators, \emph{\textit{Q} control} \cite{sul00} and \emph{dynamic PID} \cite{kod06} have been described in the literature to mitigate the effect of probe losses.
Even if both manage to increase the accuracy of the scans, they do not account for other imaging artefacts that are instead addressed by the improved controllers we propose in Section \ref{sec:Improved AFM controllers}.
For the sake of completeness, we explain both of them briefly below.

 \subsection{Q control}
 \label{subsec:Q control}

Normally, as soon as a probe loss occurs, the cantilever is oscillating away from the sample with amplitude $A \approx A_{\mathrm{r}}$, therefore $e_A \approx 0$.
The faster the error increases in absolute value, the faster the feedback controller regulating $b$ can act to recover from probe loss.
The cascade of the cantilever oscillating in free air and the demodulator can be seen as a first order system, in which the input is the dither piezo driving amplitude $D$, and the state and output is the cantilever oscillation amplitude $A$ \cite{fai13}.
This system has time constant $\tau_A = 2Q/\omega_{\mathrm{n}}$.
Therefore, the cantilever can be made more reactive by reducing the effective quality factor $Q$, which, in turn, may be achieved by changing the input from the dither piezo in (\ref{eq:dither_piezo}) to
\begin{equation}
u(t) = D\sin (\omega_{\mathrm{d}} t) - K_Q x_2.
\end{equation}
In so doing, the new effective $Q$, called $Q'$, becomes
\begin{equation}
\frac{\omega_{\mathrm{n}}}{Q} + K_Q = \frac{\omega_{\mathrm{n}}}{Q'}
\quad \Rightarrow \quad
Q' = \left( Q + \frac{K_Q}{\omega_{\mathrm{n}}} \right)^{-1}.
\end{equation}
Thus, given a desired $Q'$, the gain of the \textit{Q} control law must be chosen as
\begin{equation}
K_Q = \omega_{\mathrm{n}} \left( \frac{1}{Q'} - \frac{1}{Q} \right) .
\end{equation}
Furthermore, since $A_{\mathrm{f}}$ depends on $Q$ (see (\ref{eq:A_f})), to avoid changing $A_{\mathrm{f}}$, a new value $D'$ must be set as
\begin{equation}
D' (Q') = A_{\mathrm{f}} \left| \omega_{\mathrm{n}}^2 + \cfrac{\omega_{\mathrm{n}}}{Q'} i \omega_{\mathrm{d}} + ( i \omega_{\mathrm{d}} )^2 \right|.
\end{equation}
Note that \textit{Q} control requires the velocity signal $x_2$ to be measurable, which is true if the position signal is measured via a laser Doppler vibrometer \cite{pay12hig}.
If this is not the case, a good estimate of velocity is available applying a phase shift and a normalization to the position signal or using the so-called \emph{resonant controller}, as described in \cite{fai13}.
For a more recent implementation see \cite{rup16}.


 \subsection{Dynamic PID}
 \label{subsec:Dynamic PID}

The aim of the dynamic PID control is the same as that of \textit{Q} control, i.e.~reducing the negative effect of probe losses, but the way it achieves this goal is different; so much so that the two schemes can (and often are) used together \cite{fai13}.
Dynamic PID addresses the problem of the error $e_A$ saturating to the value $A_{\mathrm{r}}-A_{\mathrm{f}}$.
To overcome this issue, the control law (\ref{eq:b}) is modified as explained below.
Specifically, occurrence of a probe loss is inferred by inspecting the oscillation amplitude $A$: if it exceeds a threshold $A_{\mathrm{t}} > A_{\mathrm{r}}$, this means that the cantilever oscillation amplitude is not being limited by proximity with the sample surface and thus a probe loss has occurred.
When this happens, part of the error is multiplied by a gain $K_{\mathrm{s}}$:
\begin{equation}
b = \begin{dcases}
\mathrm{PID}(A_{\mathrm{r}} - A),  &  A \le A_{\mathrm{t}}  \\
\mathrm{PID}[(A_{\mathrm{r}} - A_{\mathrm{t}}) + K_{\mathrm{s}} (A_{\mathrm{t}} - A)],  &  A > A_{\mathrm{t}}
\end{dcases} ,
\end{equation}
and the PID control action is defined as in (\ref{eq:pid}).
Note that, when in probe loss, only the part of the error exceeding $A_{\mathrm{r}} - A_{\mathrm{t}}$ is multiplied by the gain, so that the output $b$ of the regulator remains a continuous signal.

 \subsection{Open problems and imaging artefacts}
 \label{subsec:Open problems and imaging artefacts}

\begin{figure}[t]
 \centering
 \includegraphics[max width=\columnwidth]{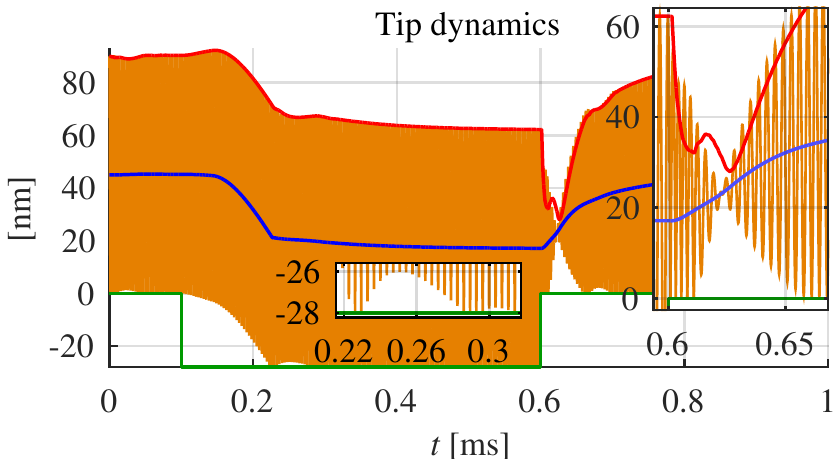}
 \caption{Scan of an ideal calibration grid, made up of a downward and an upward step, both 28 nm tall, using dynamic PID and \textit{Q} control; $A_{\mathrm{f}} = 50 \ \text{nm}$. The orange line is the absolute tip position $x_1 + b$; the red line on the top represents $A + b$; the blue line in the middle is the cantilever base height $b$; the green line on the bottom is the sample surface height $\sigma$. The inset on the left shows recovery in detail, while the one on the right illustrates recoil. The result of these phenomena is that the white space between the orange envelope and the green line is erroneously considered part of the sample surface in the measurement process.}
 \label{fig:artefacts_dynamic_pid}
 \end{figure}

While probe loss is extensively studied in the literature (e.g.,~\cite{and08}), there exist two other subtler image artefacts that can equally deteriorate image quality but are less investigated: we shall term them as \emph{recoil} and \emph{recovery}.
Both are illustrated in Figure \ref{fig:artefacts_dynamic_pid}, which shows the result of a numerical simulation that includes both \textit{Q} control and dynamic PID.

\textit{Recoil} happens when the sample to be imaged presents a steep upward step (see Figure \ref{fig:artefacts_dynamic_pid}, $t \approx 0.6 \text{\ ms}$).
In that case, the cantilever-sample separation $b-\sigma$ suddenly decreases and the interaction forces increase; as a consequence, the oscillation amplitude $A$ decreases quickly to a value smaller than $b - \sigma$ and the oscillating cantilever loses contact with the sample.
During this time, the feedback controller is ineffective, because the value of $A$ is not representative of the actual distance between the cantilever and the sample.
When the undershoot of $A$ is finished, $A$ returns to depend solely on the current cantilever-sample distance $b-\sigma$ and recoil is completed.
The effect of a recoil on surface estimation is an image artefact shaped like a bump, because $\hat \sigma = b - A$ is larger if $A$ is smaller, during the undershoot.

\textit{Recovery} occurs after dynamic PID has brought back the cantilever close to the surface, following a probe loss. In this situation there is a very short time in which the regulator keeps decreasing $b$, even if the cantilever is close to the sample surface; this delay is caused by the finite bandwidth of the feedback controller and the demodulator.
As a result (similarly to recoil) the interaction forces cause the oscillation amplitude to decrease to a value smaller than $b-\sigma$, and the cantilever detaches from the sample surface until the undershoot on $A$ finishes.
The phenomenon is observable in Figure \ref{fig:artefacts_dynamic_pid} for $t \approx 0.25 \text{ ms}$, and the artefact it generates is a false bump, just as for recoil.

\begin{figure}[t]
 \centering
 \includegraphics[max width=\columnwidth]{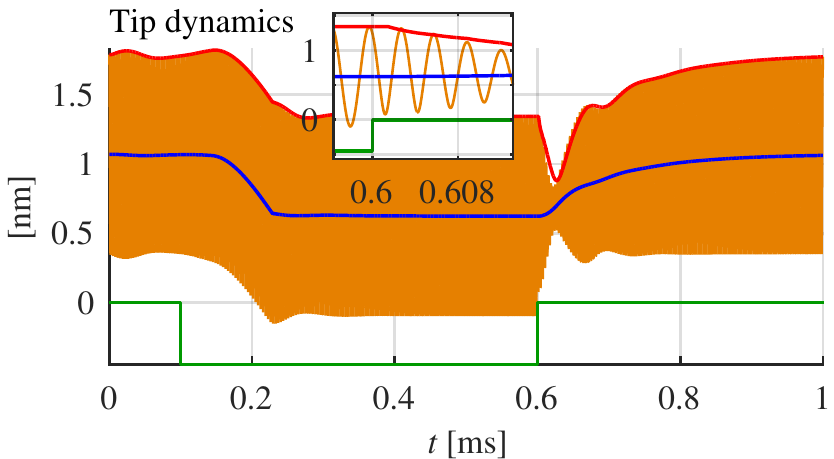}
 \caption{Scan of an ideal calibration grid with 0.448 nm tall steps, using dynamic PID and \emph{Q} control; $A_{\mathrm{f}} = 0.8 \ \text{nm}$. The inset shows that reset law (\ref{eq:hybrid_system_3}) is never triggered; despite this both recovery and recoil phenomena can be seen.}
 \label{fig:artefacts_dynamic_pid_small_A}
 \end{figure}

Note that neither event is caused by the presence of the reset law (\ref{eq:hybrid_system_3}).
In fact, as Figure \ref{fig:artefacts_dynamic_pid_small_A} shows, the phenomena can happen even when $A_{\mathrm{f}}$ is so small that the reset law is never triggered.

\section{Improved AFM controllers}
\label{sec:Improved AFM controllers}

In this section, three new proposed control schemes are illustrated.
Firstly, a hybrid PID strategy is used to deal with recovery and recoil, allowing for higher image quality.
Secondly, a scan speed regulator automatically adapts the scan velocity to the features of the sample, resulting in smaller scan time and greater accuracy.
Lastly, a predictive controller achieves the same result by forecasting the features of the specimen exploiting information deriving from already scanned portions.

 \subsection{Hybrid PID}
 \label{subsec:Hybrid PID}

 \begin{figure}[t]
 \centering
 \includegraphics[max width=\columnwidth]{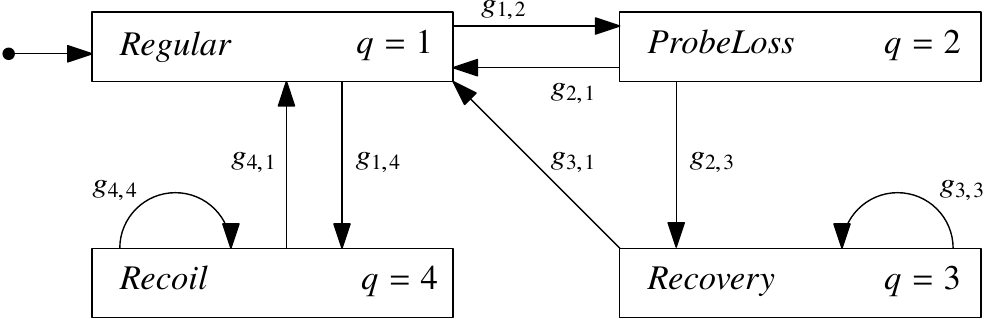}
 \caption{Hybrid PID scheme. The arrow starting from a black dot represents the initial state. Guards $g_{i,j}$ are described in Table \ref{tab:hybrid_pid_guards}.} 
 \label{fig:hybrid_pid_scheme_words}
 \end{figure}

\begin{table}[t]
\renewcommand{\arraystretch}{\rowheight}
\caption{Guards for the transitions in hybrid PID. Braces contain actions performed during the transitions.}
\label{tab:hybrid_pid_guards}
\begin{center}
\begin{tabular}{@{}lll@{}}
\toprule
Name  &  Condition, \{action\}  & Type \\
\midrule
$g_{1,2}$ & $A \ge A_{\mathrm{t}}^+$ & threshold \\
$g_{2,1}$ & $A \le A_{\mathrm{t}}^-$ & threshold \\
$g_{2,3}$ & $\dot{A} < \alpha_{\mathrm{t}}, \; \{ \rho \rightarrow \text{false}; t_0 = t \} $ & impact \\
$g_{3,3}$ & $\dot{A} > 0 \wedge \rho = \text{false}, \; \{ \rho \rightarrow \text{true} \} $  & wait \\
$g_{3,1}$ & $( \dot{A} < 0 \wedge \rho = \text{true} ) \vee (t - t_0 \ge K_\tau \tau_A )$ & impact or timeout \\
$g_{1,4}$ & $A \le A_{\mathrm{t,RL}}, \; \{ \rho \rightarrow \text{false}; t_0 = t \} $ & threshold \\
$g_{4,4}$ & $\dot{A} > 0 \wedge \rho = \text{false}, \; \{ \rho \rightarrow \text{true} \} $ & wait \\
$g_{4,1}$ & $( \dot{A} < \alpha_{\mathrm{t}} \wedge \rho = \text{true} ) \vee (t - t_0 \ge K_\tau \tau_A )$ & impact or timeout \\
\bottomrule
\end{tabular}
\end{center}
\end{table}

To address the problems caused by probe losses, recoveries and recoils, we propose a \emph{hybrid PID} strategy which combines the use of the \textit{z}-axis piezo --- which varies $b$ --- with the dither piezo --- which causes the oscillation of the cantilever. 
As Figure \ref{fig:hybrid_pid_scheme_words} shows, the controller has 4 possible modes, of which only one is active at any time; the discrete variable $q \in \{ 1, 2, 3, 4 \}$ identifies the current mode.
In all modes, the \textit{z}-axis piezo output is determined by the control law
\begin{equation}
b = \begin{dcases}
\mathrm{PID}(A_{\mathrm{r}} - A),  &  A \le A_{\mathrm{t}}^+  \\
\mathrm{PID}[(A_{\mathrm{r}} - A_{\mathrm{t}}^+) + K_{\mathrm{s}}^q (A_{\mathrm{t}}^+ - A)],  &  A > A_{\mathrm{t}}^+
\end{dcases} ,
\end{equation}
while the dither piezo output is chosen as
\begin{equation}
u = D^q \sin (\omega_{\mathrm{d}} t) - K_Q^q x_2,
\end{equation}
where variables $K_{\mathrm{s}}^q$, $D^q$ and $K_Q^q$ depend on the current mode.
As depicted in Figure \ref{fig:hybrid_pid_scheme_words}, normally --- i.e.~in absence of probe loss, recovery and recoil --- \emph{Regular} ($q=1$) is the active mode.
If, at a certain point, a probe loss (with subsequent recovery) or a recoil are detected, the controller switches to a different mode and the behaviors of the piezos change accordingly.
Specifically,
\begin{equation}
K_{\mathrm{s}}^q = \begin{dcases}
1,  &  q = 1, 3, 4 \\
K_{\mathrm{s}},  &  q = 2 \\
\end{dcases} ;
\end{equation}
which simply means that \emph{Regular}, \emph{Recovery} and \emph{Recoil} modes ($q=1,3,4$, respectively) use a regular PID, while \emph{ProbeLoss} mode ($q=2$) employs a dynamic PID.
Also, the mode-dependent control parameters $D^q$ and $K_Q^q$ are defined to be
\begin{equation}
D^q = \begin{dcases}
D',  &  q = 1, 2 \\
A_{\mathrm{f}}  \left| \omega_{\mathrm{n}}^2 + \cfrac{\omega_{\mathrm{n}}}{Q_{\mathrm{PL}}(A)}i \omega_{\mathrm{d}} + \left( i \omega_{\mathrm{d}} \right)^2  \right|,  &  q = 3 \\
A_{\mathrm{f}}  \left| \omega_{\mathrm{n}}^2 + \cfrac{\omega_{\mathrm{n}}}{Q_{\mathrm{RL}}(A)}i \omega_{\mathrm{d}} + \left( i \omega_{\mathrm{d}} \right)^2  \right|,  &  q = 4
\end{dcases} ,
\end{equation}
\begin{equation}
K_Q^q = \begin{dcases}
K_Q,  &  q = 1, 2 \\
\omega_{\mathrm{n}} \left( \frac{1}{{Q_{\mathrm{PL}}(A)}} - \frac{1}{Q} \right),  &  q = 3 \\
\omega_{\mathrm{n}} \left( \frac{1}{{Q_{\mathrm{RL}}(A)}} - \frac{1}{Q} \right),  &  q = 4
\end{dcases} ,
\end{equation}
with the probe loss ($\mathrm{PL}$) and recoil ($\mathrm{RL}$) $Q$ values
\begin{align}
Q_{\mathrm{PL}}(A) & = Q' - \Delta Q_{\mathrm{PL}} \min \left\{ \left| \frac{A_{\mathrm{r}} - A}{A_{\mathrm{r}} - A_{\mathrm{f}}} \right|, 1 \right\} , \\
Q_{\mathrm{RL}}(A) & = Q' - \Delta Q_{\mathrm{RL}} \min \left\{ \left| \frac{A_{\mathrm{r}} - A}{A_{\mathrm{r}} - 0} \right|, 1 \right\} ,
\end{align}
\begin{equation}
\Delta Q_{\mathrm{PL}}, \Delta Q_{\mathrm{RL}} > 0 .
\end{equation}
That is to say, \emph{Regular} and \emph{ProbeLoss} mode utilize a regular \textit{Q} control, whereas \emph{Recovery} and \emph{Recoil} modes employ a dynamic damping mechanism, where the further $A$ is from its reference value $A_{\mathrm{r}}$, the more the cantilever is damped.
This is to rapidly extinguish the phenomenon of the undershoot of the oscillation amplitude that happens during recoveries and recoils.

The guards that govern the transitions from one mode to another are reported in Table \ref{tab:hybrid_pid_guards} and Figure \ref{fig:hybrid_pid_scheme_words}), and may be divided into four categories:
\begin{itemize}
\item \emph{Threshold} conditions are used to detect probe losses ($g_{1,2}: A(t) > A_{\mathrm{t}}^+$) and recoils ($g_{1,4}: A(t) < A_{\mathrm{t,RL}}$), and to exit \emph{ProbeLoss} mode ($g_{2,1}: A(t) < A_{\mathrm{t}}^-$) if a recovery is not detected immediately after probe loss.
While probe loss is associated with an excessively large oscillation amplitude, the beginning of recoil is detected when an unusually small amplitude is achieved; therefore $A_{\mathrm{t}}^+ > A_{\mathrm{r}}$ and $A_{\mathrm{t,RL}} < A_{\mathrm{r}}$.
Moreover, in order to obtain a controller which is less subject to noise on $A$, $A_{\mathrm{t}}^-$ must be selected so that $A_{\mathrm{t}}^- < A_{\mathrm{t}}^+$, creating a sort of hysteresis between \emph{Regular} and \emph{ProbeLoss} modes.
\item \emph{Impact} conditions are of the kind $\mathrm{d}A/\mathrm{d}t = \dot A < 0$, and are employed to detect the beginning and end of recovery and the end of recoil.
In fact, recovery begins after probe loss when the oscillating cantilever impacts the sample surface ($g_{2,3}$), then the cantilever briefly detaches from the sample and the phenomenon ends after a second impact with the surface ($g_{3,1}$).
Similarly, recoil terminates when the cantilever oscillating in free air impacts the sample surface ($g_{4,1}$).
A threshold $\alpha_{\mathrm{t}}$ is included in $g_{2,3}$ and $g_{4,1}$ to account for signal noise on $\mathrm{d}A/\mathrm{d}t$, whereas it is absent in $g_{3,1}$, where the impact is expected to happen gently and $\mathrm{d}A/\mathrm{d}t$ is monotone;
\item \emph{Wait} conditions are used in \emph{Recovery} ($g_{3,3}$) and \emph{Recoil} ($g_{4,4}$) modes with the purpose of waiting for a change in the sign of $\mathrm{d}A/\mathrm{d}t$, in order to allow for a correct detection of impacts; the completion of such event is signalled by the Boolean variable $\rho$;
\item \emph{Timeout} conditions are set along with the impact conditions in $g_{3,1}$ and $g_{4,1}$ for those cases where impacts are not detected.
\end{itemize}


\subsection{Scan speed regulator}
\label{subsec:Scan speed regulator}

We present next an additional control scheme aimed at reducing scan time, which can be achieved by employing at all times the largest scan speed that allows for a correct imaging.
Ideally, the best way to accomplish this would be to adjust the scan speed $v_x$ dynamically, according to the rate of change of the sample surface, $|\mathrm{d}\sigma/\mathrm{d}t|$, so that when the latter is large (small), the former is small (large).
However, $|\mathrm{d}\sigma/\mathrm{d}t|$ is not easily measurable, therefore we propose that $v_x$ may be varied depending on the time-derivative $\mathrm{d}b/\mathrm{d}t$ of the \textit{z}-axis piezo input generated by the PID controller, since, if $|\mathrm{d}b/\mathrm{d}t|$ is large (small), $|\mathrm{d}\sigma/\mathrm{d}t|$ is likely to be large (small) as well.
Furthermore, $|\mathrm{d}\sigma/\mathrm{d}t|$ is actually a function of $v_x$, in the sense that if $v_x \rightarrow 0$, the surface height $\sigma$ does not change under the cantilever and $|\mathrm{d}\sigma/\mathrm{d}t| \rightarrow 0$ too.
Thus, $v_x$ must be set so that $|\mathrm{d}\sigma/\mathrm{d}t|$ (i.e. $|\mathrm{d}b/\mathrm{d}t|$) is kept within some acceptable ranges.
These ranges can be chosen considering that, adopting a hybrid PID strategy, the most critical values of $\mathrm{d}b/\mathrm{d}t$ are $b_{\mathrm{M,a}} = K_{\mathrm{I}} (A_{\mathrm{r}} - A_{\mathrm{t,RL}})$ and $b_{\mathrm{M,d}} = K_{\mathrm{I}} (A_{\mathrm{r}} - A_{\mathrm{t}}^+)$.
The former, $b_{\mathrm{M,a}}$ (\quotmarks{maximum ascending}), is the positive value of $\mathrm{d}b/\mathrm{d}t$ that, when reached, causes the hybrid PID to switch to \emph{Recoil} mode, whereas the latter, $b_{\mathrm{M,d}}$ (\quotmarks{maximum descending}), is the negative value of $\mathrm{d}b/\mathrm{d}t$ that causes the switch to \emph{ProbeLoss} mode.
Both should be avoided, in order not to trigger recoil or probe loss.
In light of this, a set of four parameters, $b_{\mathrm{L,a}}$, $b_{\mathrm{r,a}}$, $b_{\mathrm{L,d}}$ $b_{\mathrm{r,d}}$, have to be selected.
Specifically:
\begin{itemize}
\item $b_{\mathrm{L,a}} < b_{\mathrm{M,a}}$ (\quotmarks{limit ascending}) is the positive upper bound for $\mathrm{d}b/\mathrm{d}t$. The scan speed regulator is set so that $\mathrm{d}b/\mathrm{d}t$ is kept below $b_{\mathrm{L,a}}$, in order to ensure $\mathrm{d}b/\mathrm{d}t < b_{\mathrm{M,a}}$ at all times;
\item $b_{\mathrm{r,a}} < b_{\mathrm{L,a}}$ (\quotmarks{reference ascending}) is the positive reference value for $\mathrm{d}b/\mathrm{d}t$ attained by the regulator when $\mathrm{d}b/\mathrm{d}t > 0$;
\item $b_{\mathrm{L,a}} > b_{\mathrm{M,a}}$ (\quotmarks{limit descending}) is the negative lower bound for $\mathrm{d}b/\mathrm{d}t$, with the purpose of guaranteeing $\mathrm{d}b/\mathrm{d}t > b_{\mathrm{M,d}}$;
\item $b_{\mathrm{r,d}} > b_{\mathrm{L,d}}$ (\quotmarks{reference descending}) is the negative reference value for $\mathrm{d}b/\mathrm{d}t$ when $\mathrm{d}b/\mathrm{d}t < 0$.
\end{itemize}
The result is that the parameters are ordered as follows:
\begin{equation}
b_{\mathrm{M,d}} < b_{\mathrm{L,d}} < b_{\mathrm{r,d}} < 0 < b_{\mathrm{r,a}} < b_{\mathrm{L,a}} < b_{\mathrm{M,a}}.
\end{equation}
We propose to set scan velocity $v_x$ adaptively as the solution of the following first order piecewise-smooth adaptation law:
\begin{equation}\label{eq:dv_x}
\dot{v}_x = \begin{dcases}
 \cfrac{1}{\tau_v} \left[ - v_x + \left( V_{x,\mathrm{M}} - K_{v,\mathrm{a}} \left| \cfrac{\mathrm{d}b}{\mathrm{d}t} - b_{\mathrm{r,a}} \right| \right) \right],
 & \phantom{b_{\mathrm{r,d}} \le {}} \cfrac{\mathrm{d}b}{\mathrm{d}t} > b_{\mathrm{r,a}} \\
 \cfrac{1}{\tau_v} \left[ - v_x + V_{x,\mathrm{M}} \right],
 & b_{\mathrm{r,d}} \le \cfrac{\mathrm{d}b}{\mathrm{d}t} \le b_{\mathrm{r,a}} \\
 \cfrac{1}{\tau_v} \left[ - v_x + \left( V_{x,\mathrm{M}} - K_{v,\mathrm{d}} \left| \cfrac{\mathrm{d}b}{\mathrm{d}t} - b_{\mathrm{r,d}} \right| \right) \right],
 & \phantom{b_{\mathrm{r,d}} \le {}} \cfrac{\mathrm{d}b}{\mathrm{d}t} < b_{\mathrm{r,d}}
\end{dcases}
\end{equation}
\begin{equation}
K_{v,\mathrm{a}} = \frac{V_{x,\mathrm{M}}}{ \left| b_{\mathrm{L,a}} - b_{\mathrm{r,a}} \right| }, \qquad K_{v,\mathrm{d}} = \frac{V_{x,\mathrm{M}}}{ \left| b_{\mathrm{L,d}} - b_{\mathrm{r,d}} \right| } .
\end{equation}
Here, $V_{x,\mathrm{M}}$ is the (arbitrary or physical) maximum speed of the piezo maneuvering the \textit{x}-axis and $\tau_v$ is a time constant that must be compatible with the time response of the piezo.
The difference between the three cases in (\ref{eq:dv_x}) is the input: it drives $v_x$ to the maximum value $V_{x,\mathrm{M}}$ if $\mathrm{d}b/\mathrm{d}t$ is between its reference values $b_{\mathrm{r,d}}$ and $b_{\mathrm{r,a}}$; otherwise, it reduces $v_x$ all the way down to zero as $\mathrm{d}b/\mathrm{d}t$ approaches $b_{\mathrm{L,a}}$ or $b_{\mathrm{L,d}}$.
However, since for practical reasons it is better not to arrest the piezo completely, a limit is set on minimum velocity as well, so that, at any time,
\begin{equation}
 V_{x,\mathrm{m}} \le v_x \le V_{x,\mathrm{M}} .
\end{equation}
The initial value of the scan speed, $V_{x,0}$, may be set either close to $V_{x,\mathrm{m}}$, if there is a desire to act more conservatively and privilege image accuracy, or close to $V_{x,\mathrm{M}}$, if a fast scan is the priority.  
Among the advantages of this control technique is the use of different scan speeds for the ascending and descending parts of the samples, since only the latter threaten probe loss and thus require greater care.


 \subsection{Predictive controller}
 \label{subsec:Predictive controller}
 
 \begin{figure}[t]
 \centering
 \includegraphics[max width=\columnwidth]{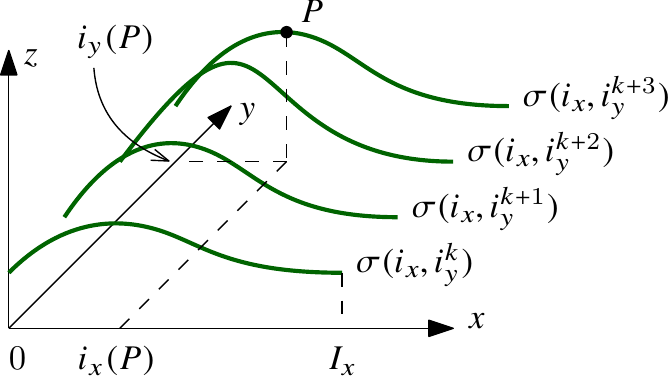}
 \caption{Representation of scan lines in the three-dimensional space.} 
 \label{fig:3d_space}
 \end{figure}
 
In the framework of the raster scan pattern, we propose to use a predictive controller to exploit information acquired from previous lines in the scan of the current one.
As shown in Figure \ref{fig:3d_space}, let
\begin{itemize}
\item $i_x(P)$ be the \textit{x} coordinate [m] of a point $P$ on the sample surface;
\item $i_y(P)$ its \textit{y} coordinate [m];
\item $I_x$ the length [m] of a scan line;
\item $I_y$ the length [m] of the sample along the \textit{y}-axis.
\end{itemize}
We suggest to extend the PID controller by adding a term to the standard PID regulator (\ref{eq:b}) of the form
\begin{equation}\label{eq:predictive_controller}
b(i_x,i_y^k) = \mathrm{PID}(A_{\mathrm{r}} - A)  + \sum\limits_{j = 1}^{M_{\mathrm{PC}}} K_{\sigma , j} \hat \sigma' (i_x, i_y^{k-j}) .
\end{equation}
Here, $M_{\mathrm{PC}}$ is the number of previous lines used, i.e.~the memory horizon of the predictive controller, $\hat \sigma' (i_x, i_y^{k-j})$ is a filtered version of the estimation of the $(k-j)$-th line and $K_{\sigma,j}$ are adaptive gains.
Converting information derived from scanned lines into a feedforward action for $b$ is straightforward, because, in a proper scan, $b$ is just a reproduction of $\sigma$, with the oscillation amplitude $A$ acting like a cushion to give the feedback controller the necessary time to adjust $b$ to $\sigma$.
Thus, in this scheme, after the first $M_{\mathrm{PC}}$ lines, the role of the PID is not to estimate $\sigma$ on a line, but just to compensate the differences between the past $M_{\mathrm{PC}}$ lines and the current one.
In Equation (\ref{eq:predictive_controller}), the sample surface estimation $\hat \sigma'$ has the \quotmarks{prime} symbol because it is actually a window-filtered version of the original, i.e.
\begin{equation}
\hat \sigma' (i_x, i_y^k) = \frac{1}{2N_{\mathrm{W}}} \int_{i_x-N_{\mathrm{W}}}^{i_x+N_{\mathrm{W}}} \hat \sigma (\xi, i_y^k) \, \mathrm{d} \xi .
\end{equation}
where it is assumed that $\hat \sigma(\xi, i_y^k) = \hat \sigma(0, i_y^k), \; \xi \in [-N_{\mathrm{W}}, 0)$ and $\hat \sigma(\xi, i_y^k) = \hat \sigma(I_x, i_y^k), \; \xi \in (I_x, I_x+N_{\mathrm{W}}]$.
This filtering is necessary because only the general shape of the scan lines is likely to recur
in the following ones.
In addition, the adaptive gains are given by
\begin{equation}
K_{\sigma , j} = \begin{dcases}
\cfrac{1}{2j} \max \left\{ \cfrac{E_\sigma - e_{\sigma , j}}{E_\sigma}, 0 \right\},  &  j \in [1, M_{\mathrm{PC}}-1]\\
\cfrac{1}{2(j - 1)} \max \left\{ \cfrac{E_\sigma - e_{\sigma , j}}{E_\sigma}, 0 \right\},  &  j = M_{\mathrm{PC}}
\end{dcases} ,
\end{equation}
where
\begin{equation}
e_{\sigma , j} \equiv \int_0^{I_x} \left| \hat \sigma' (i_x,i_y^{k-j}) - \hat \sigma' (i_x,i_y^{k-j-1}) \right| \, \mathrm{d}i_x .
\end{equation}
Note that the gains $K_{\sigma,j}$ are normalized by the factors $1/2j$ and $1/2(j-1)$, so that their sum is, at the most, unity.
Moreover, the more recent a line (smaller $j$), the higher the coefficient. 
The results of the \quotmarks{max} operations span from 0 to 1.
In particular, when $e_{\sigma,j}$, which represents how much a line is different from the previous one, is equal to or greater than a threshold $E_\sigma$, the result is 0.
Hence, the line is too different from the previous one to be used as a predictive tool.
Conversely, if $e_{\sigma,j}$ is small, the result of the \quotmarks{max} operation is close to 1, indicating that the line is adequate for a predictive use.


\section{Numerical validation} 
\label{sec:Numerical validation}

 \subsection{Settings and Samples}
 \label{subsec:Settings and Samples}

\begin{figure*}[t]
\centering
  \subfloat[]{\includegraphics[max width=0.32\textwidth]{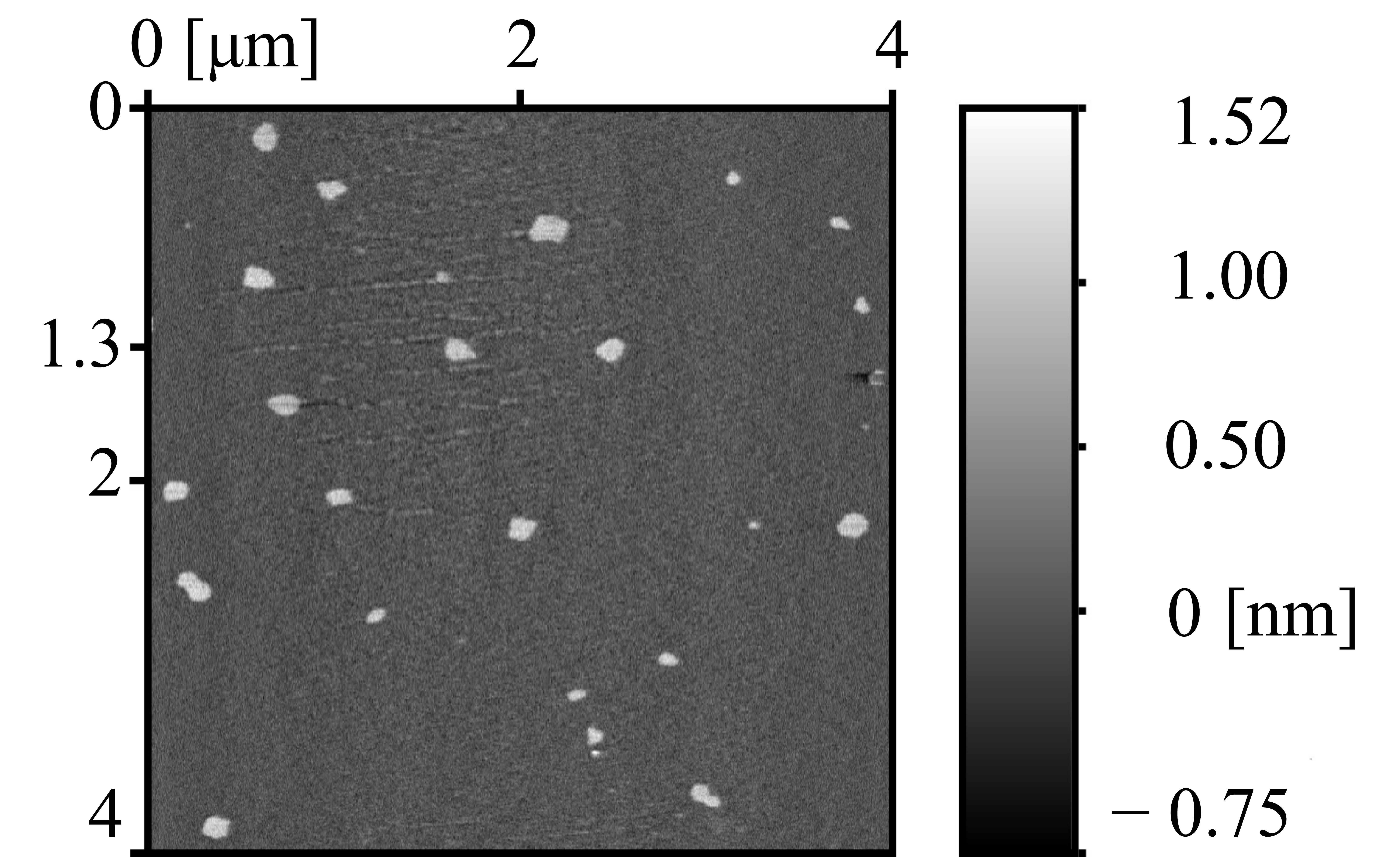}
   \label{fig:titanium_disulfide_original}}
  \subfloat[]{\includegraphics[max width=0.32\textwidth]{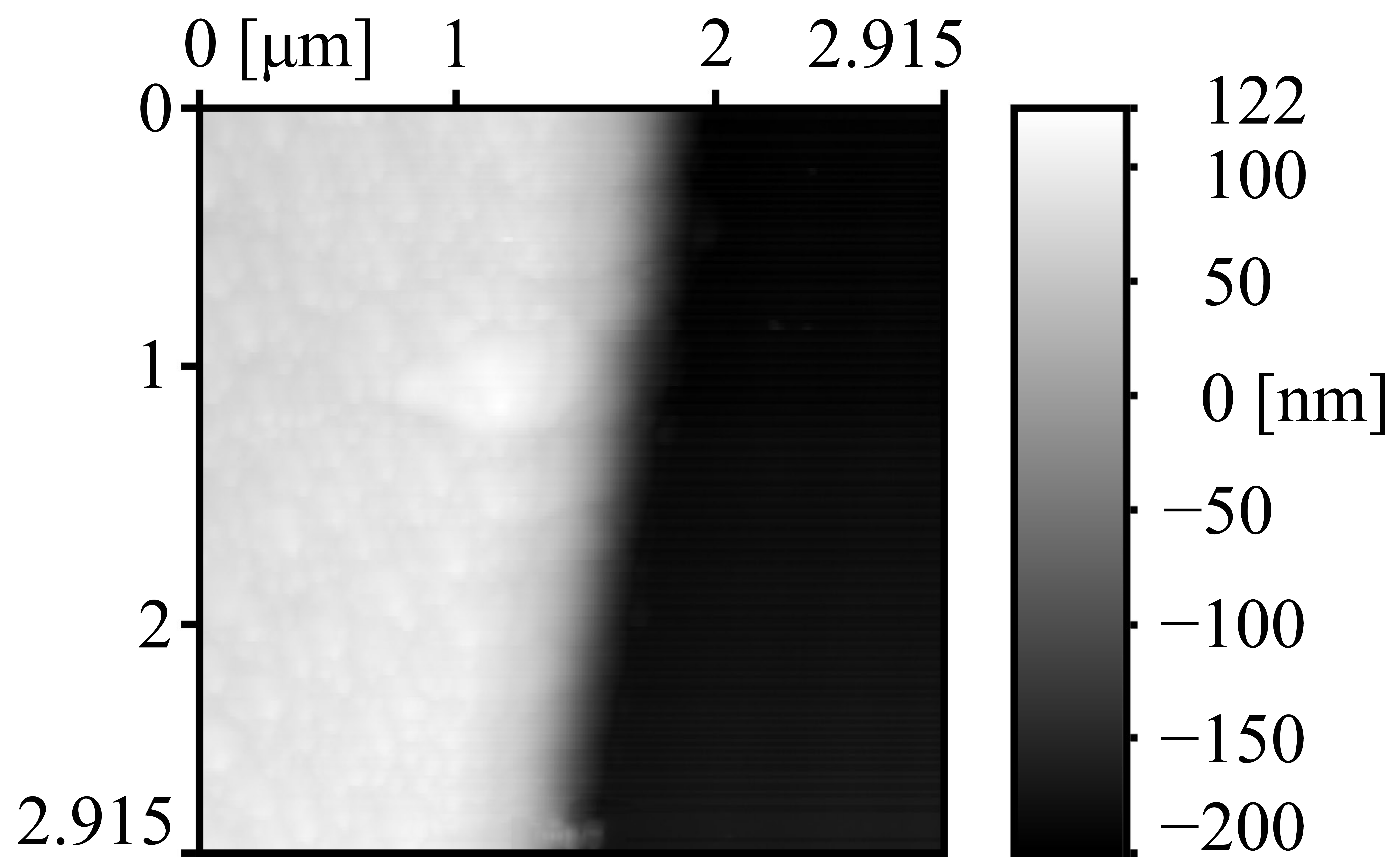}
   \label{fig:calibration_grid_original}}
  \subfloat[]{\includegraphics[max width=0.32\textwidth]{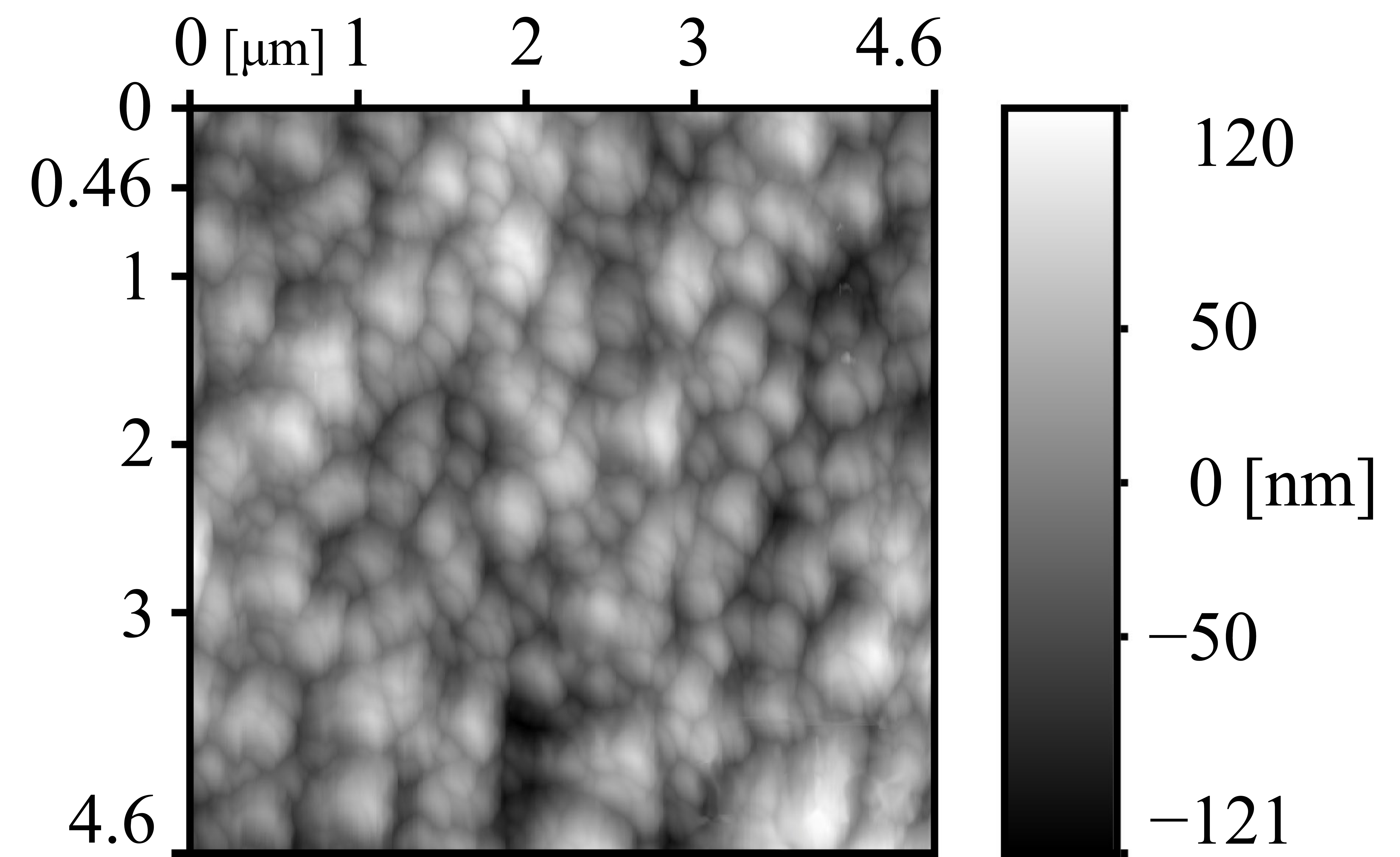}
   \label{fig:uranium_oxide_original}}
  \caption{(a) Sample of titanium disulfide; (b) calibration grid; (c) sample of uranium oxide. All the images are AFM scans of real samples.} 
  \label{fig:real_samples}  
\end{figure*}

To validate the new control strategies, we make the following assumptions regarding the AFM:
\begin{itemize}
\item The dynamics of the dither piezo are much faster than that of the system, i.e.~the largest time constant of the former is significantly smaller than $1/\omega_{\mathrm{n}}$;
\item The \textit{z}-axis piezo can be modeled as a second order system \cite{kod05}, with gain $K_{\mathrm{zp}}$, natural frequency $\omega_{\mathrm{zp}}$ and quality factor $Q_{\mathrm{zp}}$ (see Table \ref{tab:afm_parameters}).
\item \textit{Q} control is always employed and tip velocity $x_2$ is assumed measurable.
\end{itemize}
Validation will be performed on five samples: two ideal, purely numerical ones, and three real ones, previously acquired with another AFM.
These are:
\begin{itemize}
\item An ideal calibration grid, with 28 nm tall steps and a spatial period of 1 \textmu m, with each period having one downward and one upward step;
\item A real titanium disulfide sample (see Figure \ref{fig:titanium_disulfide_original}); 
\item An ideal quasi-sinusoidal sample, which is the sum of a sine having a spatial period of 4 \textmu m and an amplitude of 80 nm and a triangular waveform having amplitude and period each a tenth of those of the sine;
\item A real calibration grid sample (see Figure \ref{fig:calibration_grid_original});
\item A real uranium oxide sample (see Figure \ref{fig:uranium_oxide_original}).
\end{itemize}
All simulations were run in Matlab Simulink \cite{matlab}, using Stateflow toolbox that uses an event-driven solver to simulate the reset law (\ref{eq:hybrid_system_3}) correctly. This is coupled with a variable-step Dormand-Prince (ode45) solver, with maximum step size 10\textsuperscript{--7}, minimum step size 10\textsuperscript{--13} and relative tolerance 10\textsuperscript{--4}.
In addition, all parameters which are not expressed explicitly are taken from Table \ref{tab:afm_parameters}, unless stated otherwise.

 \subsection{Validation of hybrid PID}
 \label{subsec:Validation of hybrid PID}

 \begin{figure}[t]
 \centering
 \includegraphics[max width=\columnwidth]{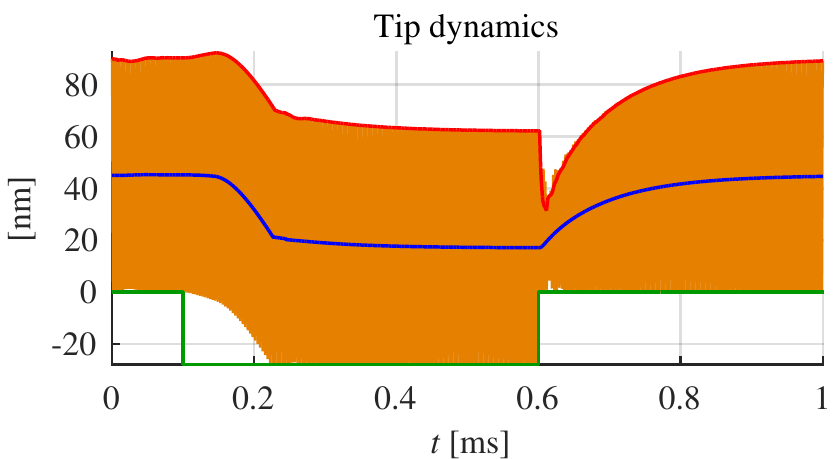}
 \caption{Scan of the ideal calibration grid with hybrid PID; $A_{\mathrm{f}} = 50 \text{ nm}$.} 
 \label{fig:hybrid_pid}
 \end{figure}

Figure \ref{fig:hybrid_pid} represents the scan of the ideal calibration grid, performed with a hybrid PID.
Compare it with Figure \ref{fig:artefacts_dynamic_pid}, where the classical dynamic PID is used on the same sample: in the former, the bump at time $t \approx 0.25 \text{ ms}$, associated with the recovery phenomenon, has practically disappeared; also, recoil decays much faster when employing the hybrid PID.
Table \ref{tab:results_hybrid_pid} reports the results of four different cases of scans of a 10-periods-long ideal calibration grid.
In the table, the variable $e_{\sigma, \mathrm{max}}^\mathrm{RV}$ indicates the maximum height of the bump observed during a recovery.
The comparison between the third rows in the first two sub-tables shows that the hybrid PID reduces the root mean square value of $e_{\sigma, \mathrm{max}}^\mathrm{RV}$ by 58.9\%.
Furthermore, the fact that the impact velocity $v_{\mathrm{i}}$ --- i.e.~the value of $x_2$ when the reset law is triggered --- does not increase points out that the new controller achieves this result without increasing the effect of the interaction forces.
The third case shows that a hybrid PID that uses \emph{Recoil} mode gives an error 6.5\% smaller than that of an hypothetical hybrid PID that does not employ it.
However, if the error is computed only during recoils, where the mode is active, the error reduction is about 20\%.
Finally, a similar result is represented in the fourth case, with noise on the position signal $x_1$, having a magnitude that is 1\% that of $A_{\mathrm{f}}$ and $\alpha_{\mathrm{t}} = - 600 A_{\mathrm{f}}$ (while in absence of noise $\alpha_{\mathrm{t}} = - 400 A_{\mathrm{f}}$).

\begin{table}[t]
\renewcommand{\arraystretch}{\rowheight}
\caption{Results of scans of a 10-period-long ideal calibration grid. $e_\sigma \mathrm{\ [nm]}$ is the estimation error; $v_{\mathrm{i}} \mathrm{\ [mm/s]}$ is the impact velocity, i.e.~when (\ref{eq:hybrid_system_3}) triggers; $e_{\sigma, \mathrm{max}}^\mathrm{RV} \mathrm{\ [nm]}$ is the height of the bump in a recovery. RMS is the root mean square value and SD is the standard deviation.}
\label{tab:results_hybrid_pid}
\begin{center}
\begin{tabular}{@{} l l S[table-format=2.2] S[table-format=2.2] S[table-format=-2.2] @{}}
\toprule
Case  &  Variable  &  {RMS}  &  {SD}  &  {Max}  \\
\midrule
1. \emph{Dynamic PID}  &  $e_\sigma$  &  8.27  &  8.61  &  28.09  \\
&  $v_{\mathrm{i}}$  &  11.78  &  10.12  &  -74.40  \\
&  $e_{\sigma, \mathrm{max}}^\mathrm{RV}$  &  2.04  &  0.77  &  3.33  \\
\hline
2. \emph{Hybrid PID w/o \emph{Recoil} mode}  &  $e_\sigma$  &  8.26  &  8.64  &  28.08  \\
&  $v_{\mathrm{i}}$  &  11.60  &  9.94  &  -74.34  \\
&  $e_{\sigma, \mathrm{max}}^\mathrm{RV}$  &  0.84  &  0.48  &  1.69  \\
\hline
3. \emph{Hybrid PID}  &  $e_\sigma$  &  7.72  &  8.21  &  28.09  \\
&  $v_{\mathrm{i}}$  &  11.38  &  9.68  &  -74.48  \\
&  $e_{\sigma, \mathrm{max}}^\mathrm{RV}$  &  1.33  &  0.65  &  2.12  \\
\hline
4. \emph{Hybrid PID with noise}  &  $e_\sigma$  &  7.88  &  7.96  &  28.18  \\
&  $v_{\mathrm{i}}$  &  10.07  &  8.69  &  -74.33  \\
&  $e_{\sigma, \mathrm{max}}^\mathrm{RV}$  &  0.92  &  0.30  &  1.35  \\
\bottomrule
\end{tabular} 
\end{center}
\end{table}

\begin{figure}[t]
\centering
  \subfloat[]{\includegraphics[max width=\columnwidth]{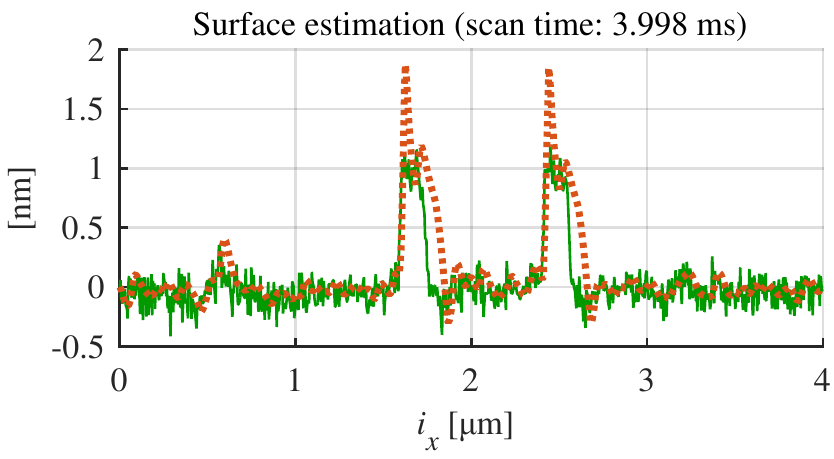}
   \label{fig:titanium_disulfide_dynamic_pid}}\\
  \subfloat[]{\includegraphics[max width=\columnwidth]{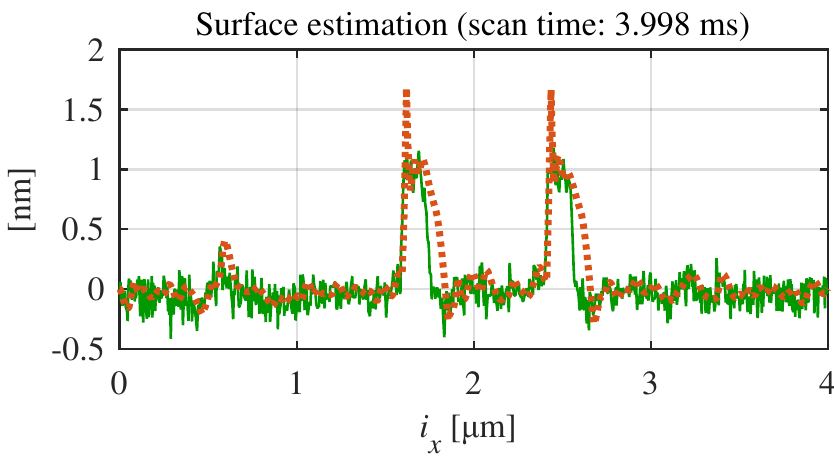}
   \label{fig:titanium_disulfide_hybrid_pid}}
  \caption{Surface estimations of the titanium disulfide sample with (a) dynamic PID and (b) hybrid PID; scan line $i_y = 1.3 \text{ \textmu m}$, $A_{\mathrm{f}} = 2 \text{ nm}$. The green solid line is the real sample surface height $\sigma$ and the orange dotted line is the estimated sample surface height $\hat \sigma$.} 
  \label{fig:titanium_disulfide}  
\end{figure}

Figures \ref{fig:titanium_disulfide_dynamic_pid} and \ref{fig:titanium_disulfide_hybrid_pid} report the surface estimations of the titanium disulfide sample on the scan line corresponding to $i_y = 1.3 \text{ \textmu m}$ when using the dynamic PID and the hybrid PID, respectively.
Table \ref{tab:results_titanium_disulfide} reports quantitative findings, showing that the root mean square error decreases by 18.2\%, when using the new scheme.

\begin{table}[t]
\renewcommand{\arraystretch}{\rowheight}
\caption{Results of scans of the titanium disulfide sample.}
\label{tab:results_titanium_disulfide}
\begin{center}
\begin{tabular}{@{} l l S[table-format=1.2] S[table-format=1.2] S[table-format=-1.2] @{}}
\toprule
Case  &  Variable  &  {RMS}  &  {SD}  &  {Max}  \\ 
\midrule
1. \emph{Dynamic PID}  &  $e_\sigma$  &  0.22  &  0.22  &  0.98  \\
&  $v_{\mathrm{i}}$  &  0.76  &  0.24  &  -1.10  \\
\hline
2. \emph{Hybrid PID}  &  $e_\sigma$  &  0.18  &  0.19  &  0.86  \\
&  $v_{\mathrm{i}}$  &  0.76  &   0.24  &  -1.09  \\
\bottomrule
\end{tabular}
\end{center}
\end{table}

 \subsection{Validation of scan speed regulator} 
  \label{subsec:Validation of scan speed regulator}

\begin{figure}[t]
\centering
  \subfloat[]{\includegraphics[max width=\columnwidth]{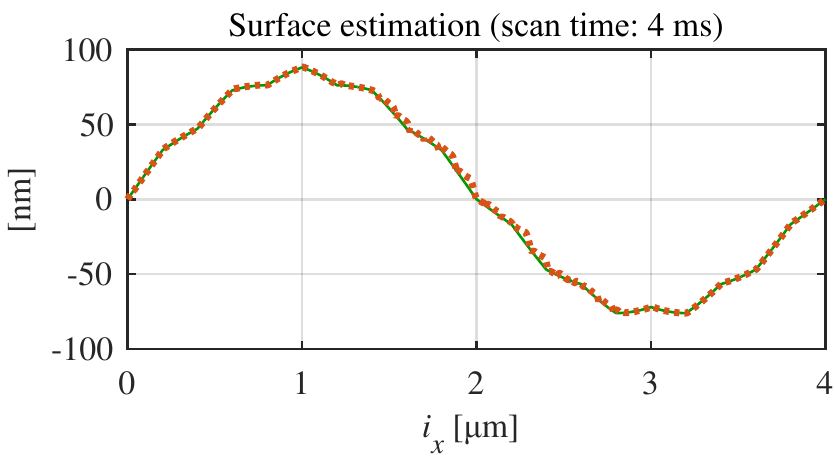}
   \label{fig:sinusoidal_sample}}\\
  \subfloat[]{\includegraphics[max width=\columnwidth]{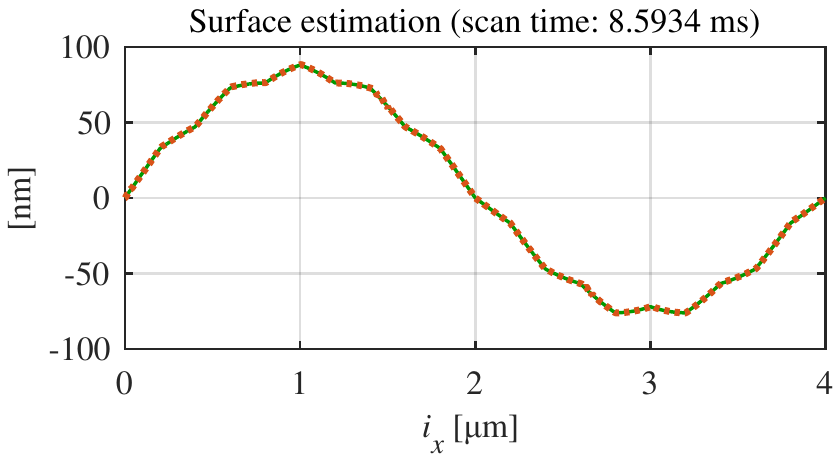}
   \label{fig:sinusoidal_sample_speed_regulator}}
  \caption{Surface estimations of the sinusoidal sample with (a) hybrid PID ($v_x = 1 \text{ mm/s}$) and with (b) hybrid PID and scan speed regulator; $A_{\mathrm{f}} = 50 \text{ nm}$.} 
  \label{fig:sinusoidal_samples}  
\end{figure}

\begin{figure}[t]
\centering
  \subfloat[]{\includegraphics[max width=\columnwidth]{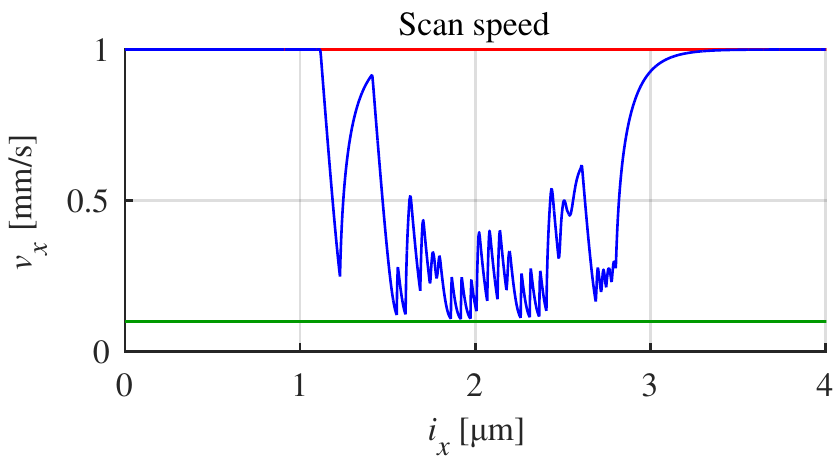}
   \label{fig:sinusoidal_sample_v_x}}\\
  \subfloat[]{\includegraphics[max width=\columnwidth]{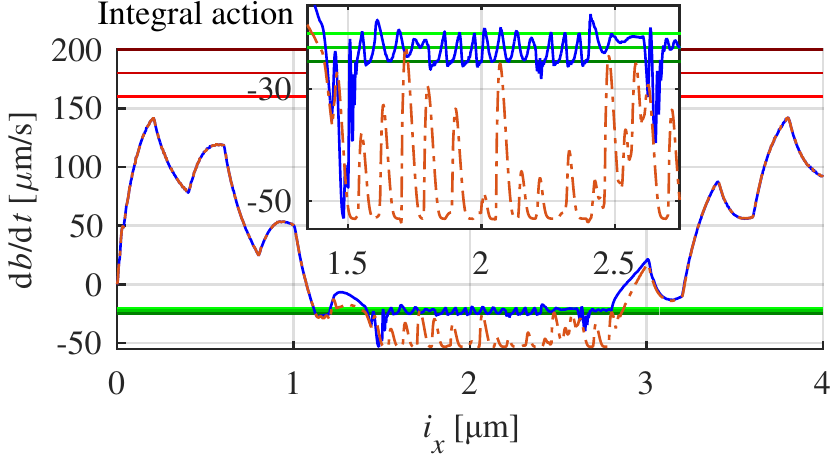}
   \label{fig:sinusoidal_sample_db}}
  \caption{(a) Scan speed $v_x$ relative to the scan in Figure \ref{fig:sinusoidal_sample_speed_regulator}. The blue line is $v_x$, the red line on the top is $V_{x,\mathrm{M}}$ and the green line on the bottom is $V_{x,\mathrm{m}}$.   
  (b) Integral action $\mathrm{d}b/\mathrm{d}t$ relative to the scans in Figures \ref{fig:sinusoidal_sample} and \ref{fig:sinusoidal_sample_speed_regulator}. The three lines on the top in shades of red are $b_{\mathrm{M,a}}$, $b_{\mathrm{L,a}}$ and $b_{\mathrm{r,a}}$, recalling that $b_{\mathrm{M,a}} > b_{\mathrm{L,a}} > b_{\mathrm{r,a}}$.   
The three lines on the bottom in shades of green are $b_{\mathrm{M,d}}$, $b_{\mathrm{L,d}}$ and $b_{\mathrm{r,d}}$, with $b_{\mathrm{M,d}} < b_{\mathrm{L,d}} < b_{\mathrm{r,d}}$.
The dashed orange line is the integral action relative the scan in Figure \ref{fig:sinusoidal_sample} (w/o speed regulator), whereas the blue solid line is the integral action relative to the scan in Figure \ref{fig:sinusoidal_sample_speed_regulator} (w/ speed regulator).
The inset shows a detail corresponding to the descending part of the sample.} 
 \label{fig:sinusoidal_sample_v_x_db}  
\end{figure}

When scanning the ideal quasi-sinusoidal sample with constant scan speed $v_x = 1 \text{ mm/s}$ and using the hybrid PID, the AFM is not able to image the sample properly and probe losses happen during the descending part of the surface, as shown in Figure \ref{fig:sinusoidal_sample}.
Instead, a nearly perfect scan is achieved when adding the scan speed regulator, with $V_{x,\mathrm{M}} = V_{x,0} = 1 \text{ mm/s}$ and $V_{x,\mathrm{m}} = V_{x,0}/10$, as depicted in Figure \ref{fig:sinusoidal_sample_speed_regulator}.
The comparison between the sub-tables in Table \ref{tab:results_sinusoidal_sample} shows that, when using the scan speed regulator, the root mean square error decreases by 86\%.
To obtain the same level of accuracy without the scan speed regulator, it would be necessary to reduce the scan speed to $v_x = 0.421 \text{ mm/s}$, as in case 3, having however the scan time increased by 10.6\% with respect to case 2.
For the sake of completeness, Figure \ref{fig:sinusoidal_sample_v_x_db} shows the evolution of $v_x$ and $\mathrm{d}b/\mathrm{d}t$ with and without scan speed regulator, corresponding to the scans shown in Figure \ref{fig:sinusoidal_samples}.

\begin{table}[t]
\renewcommand{\arraystretch}{\rowheight}
\caption{Results of scans of the sinusoidal sample. $T_\mathrm{s} \mathrm{\ [ms]}$ is the scan time.}
\label{tab:results_sinusoidal_sample}
\begin{center}
\begin{tabular}{@{} l l S[table-format=1.3] S[table-format=1.2] S[table-format=-2.2] @{}}
\toprule
Case  &  Variable  &  {RMS}  &  {SD}  &  {Max}  \\ 
\midrule
1. \emph{Hybrid PID ($v_x = 1 \text{ \rm{mm/s}}$)} &  $e_\sigma$  &  1.66  &  1.66  &  8.24  \\
&  $v_{\mathrm{i}}$  &  5.88  &  2.88  &  -19.77  \\
&  $T_\mathrm{s}$  &  3.998  &  {-}  &  {-}  \\
\hline
2. \emph{Hybrid PID and speed regulator}  &  $e_\sigma$  &  0.23  &  0.25  &  2.01 \\
&  $v_{\mathrm{i}}$  &  4.09  &  2.02  &  -19.77  \\
&  $T_\mathrm{s}$  &  8.591  &  {-}  &  {-}  \\
\hline
3. \emph{Hybrid PID ($v_x = 0.421 \text{ \rm{mm/s}}$)}  &  $e_\sigma$  &  0.24  &  0.25  &  1.79  \\
&  $v_{\mathrm{i}}$  &  4.25  &  1.91  &  -16.84  \\
&  $T_\mathrm{s}$  &  9.501  &  {-}  &  {-}  \\
\bottomrule
\end{tabular}
\end{center}
\end{table}

\begin{figure}[t]
\centering
  \subfloat[]{\includegraphics[max width=\columnwidth]{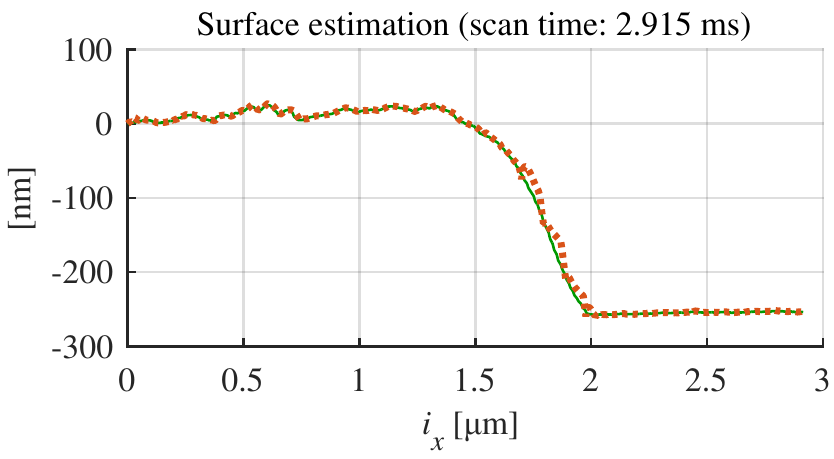}
   \label{fig:calibration_grid}}\\
  \subfloat[]{\includegraphics[max width=\columnwidth]{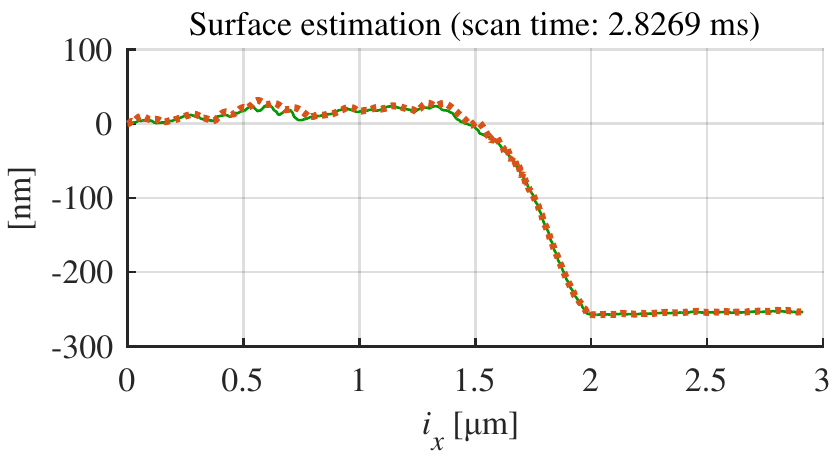}
   \label{fig:calibration_grid_speed_regulator}}
  \caption{Surface estimations of the real calibration grid using (a) the hybrid PID ($v_x = 1 \text{ mm/s}$) and (b) the hybrid PID and the scan speed regulator ($V_{x,0} = 2 \text{ mm/s}$.); scan line $i_y=0 \text{ \textmu m}$, $A_{\mathrm{f}} = 300 \text{ nm}$.} 
  \label{fig:calibration_grids}  
\end{figure}

To further validate these findings, compare the results of a scan of the first line ($i_y = 0 \text{ \textmu m}$) of the real calibration grid without the scan speed regulator, reported in Figure \ref{fig:calibration_grid}, with a scan performed while employing it, depicted in Figure \ref{fig:calibration_grid_speed_regulator}; quantitative results are in Table \ref{tab:results_calibration_grid}.
In particular, when using the scan speed regulator the error decreases by 47\% and the scan time by 3\%.
 
\begin{table}[t]
\renewcommand{\arraystretch}{\rowheight}
\caption{Results of scans of the real calibration grid.}
\label{tab:results_calibration_grid}
\begin{center}
\begin{tabular}{@{} l l S[table-format=2.3] S[table-format=2.2] S[table-format=-3.2] @{}}
\toprule
Case  &  Variable  &  {RMS}  &  {SD}  &  {Max}  \\ 
\midrule
1. \emph{Hybrid PID ($v_x = 1 \mathrm{\ mm/s}$)}  &  $e_\sigma$  &  5.43  &  5.76  &  3.26  \\
&  $v_{\mathrm{i}}$  &  29.66  &  13.86  &  -107.49  \\
&  $T_\mathrm{s}$  &  2.915  &  {-}  &  {-}  \\
\hline
2. \emph{Hybrid PID and speed} &  $e_\sigma$  &  2.88  &  2.98  &  15.05  \\
\phantom{2. }\emph{regulator ($V_{x,0} = 2 \mathrm{\ mm/s}$)}  &  $v_{\mathrm{i}}$  &  29.98  &  14.75  &  -91.30  \\
&  $T_\mathrm{s}$ &  2.827  &  {-}  &  {-}  \\
\bottomrule
\end{tabular}
\end{center}
\end{table}

 \subsection{Validation of predictive controller} 
 \label{subsec:Validation of predictive controller} 

\begin{figure}[t]
 \centering
 \includegraphics[max width=\linewidth]{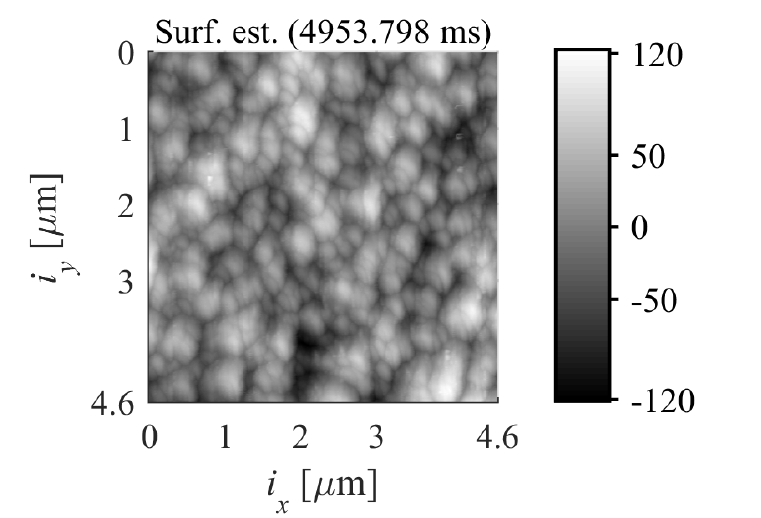}
 \caption{Surface estimation of the uranium oxide sample with hybrid PID, scan speed regulator and predictive controller; $A_{\mathrm{f}} = 200 \text{ nm}$.} 
 \label{fig:uranium_oxide_predictive_speed_regulator}
 \end{figure}

The predictive controller has been tested together with the scan speed regulator on the uranium oxide sample; Figure \ref{fig:uranium_oxide_predictive_speed_regulator} depicts a scan of the whole surface, which may be compared with the original in Figure \ref{fig:uranium_oxide_original}.
In addition, the results of a series of comparative tests are reported in Table \ref{tab:results_uranium_oxide}.
In these simulations the first 100 lines of the sample are scanned ($i_y = 0 \text{ \textmu m}$ to $i_y = 0,46 \text{ \textmu m}$), in four different configurations, given by the possible combinations of the predictive controller and the scan speed regulator.
In a scenario where the scan speed regulator is not used, adding the predictive controller reduces the error by 39.4\% (cases 1 and 2).
In contrast, when using an AFM which implements the speed regulator, the predictive controller reduces the error by 18.6\% and the scan time by 19.9\% (cases 3 and 4).
In conclusion, comparing the results given by the four configurations, the best solution is to employ the predictive controller together with the scan speed regulator, in order to have the best accuracy, reduced scan time and self-selection of scan speed.

\begin{table}[t]
\renewcommand{\arraystretch}{\rowheight}
\caption{Results of scans of uranium oxide. $T_\mathrm{s} \mathrm{\ [ms]}$ is the scan time of a single line and $T_\mathrm{s,tot} \mathrm{\ [ms]}$ is the total scan time.}
\label{tab:results_uranium_oxide}
\begin{center}
\begin{tabular}{@{} l l S[table-format=3.3] S[table-format=1.3] @{}}
\toprule
Case  &  Variable  &  {Mean}  &  {Max}  \\ 
\midrule
1. \emph{w/o predictive controller,}  &  $\mathrm{RMS}(e_\sigma)$  &  5.82  &  7.79  \\    
\phantom{1. }\emph{w/o speed regulator}  &  $\mathrm{SD}(e_\sigma)$  &  5.06  &  6.87  \\
&  $K_\sigma$  &  0.000  &  0.000  \\
&  $T_\mathrm{s}$  &  4.600  &  4.600  \\
&  $T_\mathrm{s,tot}$ &  460.000  &  {-}  \\
\hline
2. \emph{w/ predictive controller,}  &  $\mathrm{RMS}(e_\sigma)$  &  3.53  &  5.83  \\    
\phantom{2. }\emph{w/o speed regulator}  &  $\mathrm{SD}(e_\sigma)$  &  2.83  &  5.01  \\
&  $K_\sigma$  &  0.886  &  0.927  \\
&  $T_\mathrm{s}$  &  4.600  &  4.600  \\
&  $T_\mathrm{s,tot}$  &  460.000  &  {-}  \\
\hline
3. \emph{w/o predictive controller,}  &  $\mathrm{RMS}(e_\sigma)$  &  4.06  &  5.58  \\    
\phantom{3. }\emph{w speed regulator}  &  $\mathrm{STD}(e_\sigma)$  &  3.56  &  4.95  \\
&  $K_\sigma$  &  0.000  &  0.000  \\
&  $T_\mathrm{s}$  &  6.248  &  6.810  \\
&  $T_\mathrm{s,tot}$  &  624.770  &  {-}  \\
\hline
4. \emph{w/ predictive controller,}  &  $\mathrm{RMS}(e_\sigma)$  &  3.31  &  4.50  \\    
\phantom{4. }\emph{w speed regulator}  &  $\mathrm{STD}(e_\sigma)$  &  2.71  &  3.96  \\
&  $K_\sigma$  &  0.888  &  0.927  \\
&  $T_\mathrm{s}$  &  5.004  &  6.455  \\
&  $T_\mathrm{s,tot}$  &  500.359  &  {-}  \\
\bottomrule
\end{tabular}
\end{center}
\end{table}


\section{Conclusions}
\label{sec:Conclusions}

In this paper we have introduced three original controllers that achieve two fundamental goals: improving the accuracy and reducing the scan time of the intermittent contact mode atomic force microscope.
Firstly, a hybrid PID scheme was introduced which is able to deal with image artefacts such as recoils and recoveries.
Secondly, an adaptive scan speed regulator is proposed to set scan speed dynamically, depending on the characteristics of the sample surface.
As a result, scan time decreases, accuracy being equal.
Finally, a predictive controller is used to improve both the image quality and the scan time, exploiting information deriving from already scanned lines in the imaging of the current one.

Further research will focus on obtaining analytical results from the modelling of the IC-AFM together with the novel proposed controllers, in presence of simple sample surfaces such as sinusoidal or square waves, and with a piecewise linear model for the interaction forces.


\section{Acknowledgments}
\label{sec:Acknowledgements}

The authors wish to thank Mr Davide Fiore at the University of Naples Federico II for the insightful comments and discussions.
MC wishes to acknowledge the University of Naples Federico II (Italian Ministerial Decree 976, 29/12/2014 -- Art. I) for supporting his visit at the Department of Engineering Mathematics of the University of Bristol from 16/01/2016 to 20/02/2016.
MH wishes to acknowledge funding from Rete di Eccellenza MASTRI that supported his visit to Naples in 2015, and helped initiate this collaboration.






\ifCLASSOPTIONcaptionsoff
  \newpage
\fi


\bibliographystyle{IEEEtran}  
\bibliography{references}








\end{document}